# Male Earnings Volatility in LEHD before, during, and after the Great Recession

Kevin L. McKinney, *US Census Bureau*

John M. Abowd, *US Census Bureau*

January 20, 2022

**Abstract**

This paper is part of a coordinated collection of papers on prime-age male earnings volatility. Each paper produces a similar set of statistics for the same reference population using a different primary data source. Our primary data source is the Census Bureau's Longitudinal Employer-Household Dynamics (LEHD) infrastructure files. Using LEHD data from 1998 to 2016, we create a well-defined population frame to facilitate accurate estimation of temporal changes comparable to designed longitudinal samples of people. We show that earnings volatility, excluding increases during recessions, has declined over the analysis period, a finding robust to various sensitivity analyses.

Any opinions and conclusions expressed herein are those of the authors and do not represent the views of the US Census Bureau or other sponsors. All results have been reviewed to ensure that no confidential information is disclosed (DRB clearance numbers CBDRB-FY20-279, CBDRB-FY20-288, CBDRB-FY20-420, CBDRB-FY21-CED006-0023, CBDRB-FY22-089, CBDRB-FY22-CED006-0011). This research uses data from the US Census Bureau's Longitudinal Employer-Household Dynamics Program, which was partially supported by NSF grants SES-9978093, SES-0339191, and ITR-0427889; National Institute on Aging grant AG018854; and grants from the Alfred P. Sloan Foundation. The most recent version of this paper can be found at http://arxiv.org/abs/2008.00253.

## I. Introduction

This paper is part of a coordinated collection of papers on prime-age male earnings volatility. Each paper produces a similar set of statistics for the same reference population using a different primary data source. Our paper uses unemployment insurance (UI) based administrative worker annual earnings data from the Census Bureau's Longitudinal Employer Household Dynamics (LEHD) program. To put our paper in context, we note it is based on sample selection methods developed in Abowd et al. (2018; AMZ hereafter). Among other findings, AMZ demonstrated that in order to draw comparisons between administrative, designed cross-sectional, and designed longitudinal estimates of related labor market phenomena like earnings inequality and volatility, one must construct a proper, dynamic universe as a reference population. Armed with that construct, which we also employ in this paper, we show that it is essential to pay careful attention to the left tail (bottom) of the earnings distribution.

Using our consistent, dynamic population frame, we estimate earnings volatility trends for prime-age males from 1998 to 2016. Our main result is that volatility has declined over the analysis period, excluding increases during recessions. Although the level of volatility differs across various specifications, the downward trend in volatility is robust to all sensitivity analyses.

Our paper is organized as follows. Section II describes the methods we used to construct the LEHD analysis samples. Section III presents our statistical results. Section IV concludes.

## II. Data

The empirical work in this paper uses earnings information from the Longitudinal Employer-Household Dynamics (LEHD) infrastructure files, developed and maintained by the U.S. Census Bureau.[1] From this data source, we construct annual person-level earnings files covering the period 1998-2017.

In contrast with survey data sources such as the CPS, PSID, and ACS, LEHD data contain annual earnings for the virtual universe of private wage and salary workers in the United States. The essentially complete coverage of the universe enables a detailed analysis of worker earnings volatility.[2] However, it is essential to use earnings associated with only "eligible" workers—those who are part of our dynamic frame—namely, individuals with identifiers issued by the Social Security Administration, and used by the person to whom they were issued. Adopting this population frame, allows us to consistently estimate calendar-year time trends for our earnings volatility measures.

In the LEHD data infrastructure, a "job" is the statutory employment of a worker by a statutory employer as defined by the Unemployment Insurance (UI) system in a given state. Mandated reporting of UI-covered wage and salary payments between one statutory employer and one statutory employee is governed by the state's UI system. Reporting covers private employers and state and local government. There are no self-employment earnings unless the proprietor drew a salary, which, for UI earnings data, is indistinguishable from other employees.

States joined the Local Employment Dynamics federal/state partnership that supplies input data to the LEHD program at different dates.[3] When a state joined, the data custodians were asked to produce historical data for as many quarters in the past, back to 1990Q1, as could be reasonably recovered from their information storage systems. As a result, the date that a data-supplying entity joined the partnership is not the same as the first quarter in which that entity's data appear in the system. The start date for any state depends primarily on the amount of historical data the state could recover at the time it joined. This potential ignorability (in the sense of Rubin 1987 or Imbens and Rubin 2015) of the start date for a segment of the LEHD data – that is the possibility that state start dates are conceivably not related to data quality or earnings volatility – is the basis for our methods of constructing a time-series of nationally representative estimates.

---

[1] See Abowd et al. (2009) for a detailed summary of the construction of the LEHD infrastructure.
[2] One artifact of the exceptionally high job and worker coverage rates in LEHD UI data is a relatively large number of low earning workers. See Appendix Figure A1 for a comparison of the ACS, PSID, and LEHD annual earnings distributions.
[3] See Appendix Table A1 for details on state entry/exit.

Although state entry is a potential concern, AMZ show that the annual earnings distributions for the subset of states available by 1995Q1 are almost identical to the complete data annual earnings distribution. However, in this paper we are not estimating measures of earnings inequality but are instead measuring earnings volatility. The variance measures used here are especially sensitive to earnings changes at the bottom of the earnings distribution, where a relatively small change in absolute value can result in a large percentage change. Even though AMZ show the annual earnings distributions between the fifth and the ninety-fifth percentiles are not noticeably affected by state entry/exit, we need to be cautious in applying their results to this paper because earnings changes for workers in the bottom five percent of the earnings distribution potentially have an outsized impact on the results.

Although the LEHD data provide a high-quality jobs frame, individual identifier misuse complicates the time-varying many-to-one assignment of jobs to workers. Therefore, it is preferable to have a person frame that covers a known population of interest, such as all persons legally eligible to work in the United States. For our analysis, we create a frame of workers using the Census Bureau's edited version of the Social Security Administration's master SSN database (the "Census Numident"), capturing all officially reported employment-eligible workers but removing jobs associated with ineligible workers, as we elaborate below.

LEHD earnings records are reported quarterly by the employing firm. These records contain a nine-digit person identifier, typically assumed to be a Social Security Number. However, at the time the report is received by the state UI office, the nine-digit person identifier is not verified, resulting in records both with and without a valid SSN. Using the Census Numident we ascertain if each earnings record is associated with a valid SSN. Records not associated with a valid SSN may have an alternate valid person identifier such as an IRS-issued Individual Taxpayer Identification Number (ITIN); nevertheless, we can only distinguish between valid and invalid SSNs. If the SSN is valid, in addition to the UI employment history we have access to demographic characteristics, such as sex and birthdate, from the Census Numident.

Using both the Census Numident and the employment histories from the UI data, we create a "prime-age male eligible-workers" frame, including only workers in a given year that meet the following criteria: individual has a valid SSN on the Census Numident; gender of the individual is male; the year is between 1998 and 2017, inclusive; the modal age of the individual during the year is between 25 and 59, inclusive; the year is greater than the SSN year-of-issue and less than the year of death (if available); and the SSN is associated with fewer than 12 jobs during the year. An eligible worker is labeled as "active" in the labor market for a particular year when UI earnings are positive in that year and "inactive" otherwise. Inactive status is thus inferred based on the absence of positive earnings reports.[4]

---

[4] Sample sizes for our base analysis sample are shown in Appendix Table A2.

The purpose of the prime-age male eligible-workers frame is twofold. First, the Census Numident data allow us to consistently identify a set of males legally eligible to work each year, while at the same time implicitly removing earning records from our analysis sample not associated with individuals in the covered legally eligible-to-work population. However, we go a step further, removing earnings records with valid SSNs where the available data strongly suggest the SSN is not being used by the person to whom it was issued.[5] Our criteria for excluding earnings records uses three annual exclusion rules, preserving job-level earnings records during years when none of the rules are violated. The three exclusion rules are: job years with positive UI earnings prior to the SSN being issued are removed, while later years are eligible for inclusion; job years with positive reported earnings where the individual is reported dead prior to the start of the year are removed, although earlier years are eligible for inclusion; and job years where the individual has more than 12 employers are excluded, although other years are eligible for inclusion.

Our earnings measure is based on annual UI job-level earnings reports. First, we adjust nominal earnings to real earnings using the Bureau of Economic Analysis' Personal Consumption Expenditures (PCE) index, with 2010 as the base year. Second, we calculate person-level real annual earnings $e_{it}$ for each eligible male worker $i$ in each year $t$ as the sum of real earnings $v_{ijt}$ at all firms $j$ for each worker $i$ in each year $t$.

$$e_{it} = \sum_j v_{ijt} \qquad l_{it} = \ln(e_{it+1}) - \ln(e_{it}) \qquad a_{it} = \frac{(e_{it+1} - e_{it})}{(e_{it} + e_{it+1})/2}$$

To estimate earnings volatility, we create two measures of the change in annual earnings. Our first earnings volatility measure is the difference in log earnings (DLE) from the initial year $t$ to the subsequent year $t+1$ The DLE measure, $l_{it}$, is available from 1998 to 2016 for workers with positive earnings in both years. We also analyze a second measure, the arc percentage change (APC) $a_{it}$ (Dahl, DeLeire, and Schwabish, 2008, 2011; Ziliak, Hardy, and Bollinger, 2011).[6] The APC is available from 1998 to 2016 for all workers in the composite sample and unlike the DLE, the APC can be used when one of the earnings observations for the year-pair is zero.

---

[5] The use of SSNs not originally issued to the person using the SSN has been documented and studied by Brown et al. 2013, AMZ, and others.

If the change in earnings is moderate, $l_{it}$ and $a_{it}$ produce similar results. For example, when earnings decrease by less than fifty percent or increase less than one hundred percent, the relative difference between the two measures is no more than three percentage points. However, when there is an extremely large percentage change in earnings the two measures may differ substantially. For example, if a worker earns $1,000 in the first year and $50,000 in the second year of a year-pair, the difference in log earnings is over twice as large as the arc percentage change (3.91 vs 1.92). Given the sensitivity of the variance calculation to large earnings changes, the two measures may produce very different results even when presented with the same data.

## III. Results
### a. Baseline Trends

Standard practice in the literature is to separately trim initial and subsequent-year earnings using stated minimum and maximum values, thus placing an upper bound on the absolute value of the earnings change. We call this a "by-year" trim. A secondary but perhaps more important effect of trimming is that it reduces the number of very low earning workers, for whom a relatively small absolute change in earnings can result in a large percentage change. We implement trimming by year, setting the minimum at the 1st percentile and the maximum at the 99th percentile of the real earnings distribution for the actual data year. We also implement constant or "same each year" trimming using the percentiles estimated from the combined earnings distribution across all years of the composite sample. There is merit in both constant and by-year trimming. By using the same values each year, constant trimming prioritizes a worker's absolute position in the real earnings distribution, which can be especially important for workers at the bottom of the earnings distribution where the absolute level of earnings may be material to their existence. However, using time-varying trim values allows us to include the set of workers each year who are in the same relative position in the real earnings distribution.

---
[6] Note that while the APC measure is called a "percentage" change in the literature, it is defined as a proportional change.

The choice of trim values is especially important for trend analysis if there are large changes in the tails of the cross-sectional earnings distribution over time or the tails are estimated imprecisely due to small sample sizes. Trends estimated using either the constant trim or the by-year trim would be similarly affected by increases in relative volatility in the tails of the earnings distribution, however, trends estimated using a by-year percentile-based trim will likely differ from trends estimated using a constant trim because the share of workers in the tails is changing over time. For example, if the earnings distribution shifts to the left, similar to what might happen during a recession, the constant trim would remove a larger fraction of low earning workers from the analysis sample than the by-year trim, resulting in a lower estimate of earnings volatility.

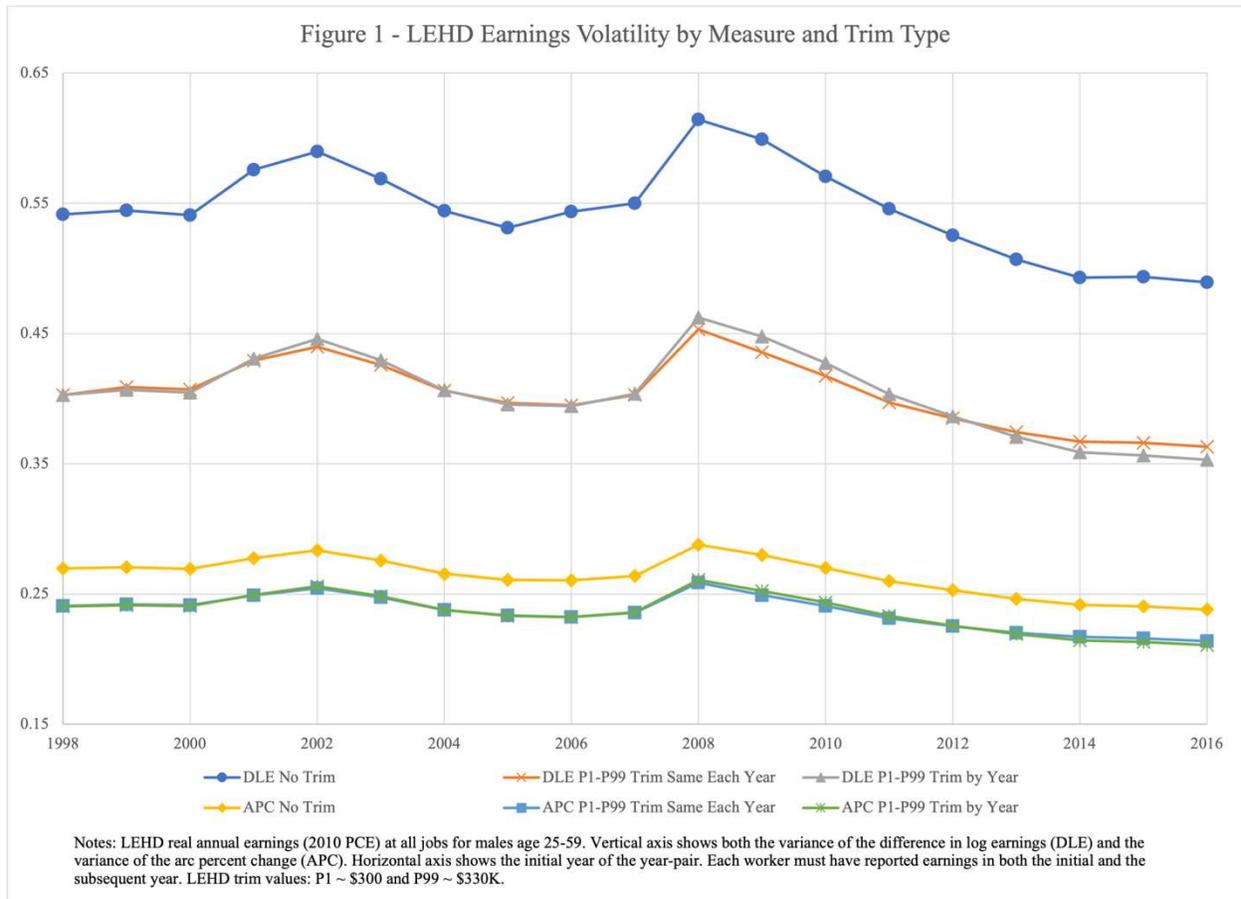

The results for both the DLE measure of volatility and the APC measure of volatility are shown in Figure 1. Three themes are evident. First, earnings volatility is counter-cyclical with increases during recessions (2000-2001, 2007-2008) and decreases during expansions. Second, ignoring the cyclicality, earnings volatility is relatively stable until 2008, when it begins a sustained decline. Third, volatility measured using the DLE is always larger than volatility measured using the APC.

The primary effect of trimming is to reduce the level of measured volatility. For example, using the DLE the untrimmed level of volatility at the 2008 peak of the great recession is about 0.62, while the trimmed values are 0.45 for the constant trim and 0.46 for the by-year trim. The substantial reduction in volatility highlights the sensitivity of the variance to the tails of the earnings distribution. The trends for the trimmed and untrimmed series are similar, although there are small, but noticeable, differences between the constant-trim and by-year trim estimates. Volatility is larger in the by-year trim series during recessions as workers face negative earnings shocks and the earnings distribution shifts to the left, resulting in a larger share of lower earnings workers in the analysis sample and higher volatility. The trends are similar for both the DLE and APC, but the effects of trimming are smaller for the APC. The APC incorporates a smooth limit to the volatility calculation, reducing the impact of the trim. For the same reason, the increase in APC volatility during recessions is also muted, irrespective of the trim.[7]

### b. Comparisons with Survey Data

Our primary analysis dataset is based on state Unemployment Insurance (UI) records. Compared with survey data sources, LEHD UI data covers a different population of jobs and workers. In this section, we document the differences in earnings volatility between the LEHD, a sample of LEHD annual earnings for workers who appear in the American Community Survey (the LEHD-ACS sample), and the PSID.

In our first exercise we use the matched LEHD-ACS sample to estimate the impact of missing earnings from federal and "self-employed in a not incorporated firm" jobs on our LEHD earnings volatility estimates. Workers that transition in and out of LEHD UI covered employment within a year-pair are especially problematic, potentially increasing our earnings volatility estimates. To create the LEHD-ACS sample, we match each LEHD worker year-pair observation with the ACS data by SSN and year, allowing us to identify records in the LEHD-ACS matched sample who are likely to have a job in employment sectors not covered in the

---

[7] Additional sensitivity analyses are found in the Appendix; Figure A2 shows the impact of state entry/exit, Figure B2 shows the impact of adjusting for age, Figure C1 compares our main result with a non-parametric measure of earnings volatility, Figure D1 shows a comparison of trim value trends, Figure D2 shows earnings volatility trends for selected other trimming approaches, and Figure E2 shows the impact of including zero earning years.

LEHD data.[8] The LEHD earnings volatility results for the LEHD-ACS sample are shown in Figure 2.

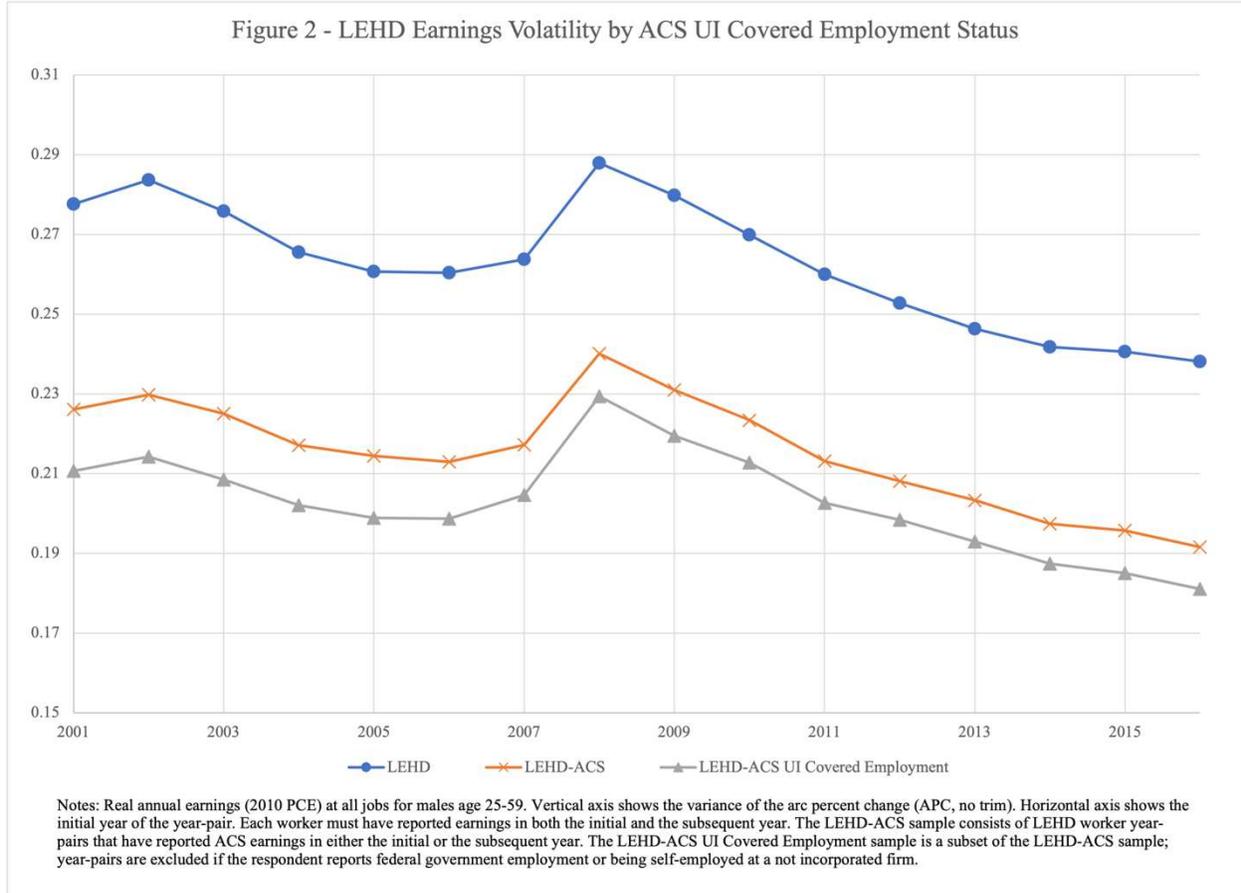

Compared with the baseline LEHD earnings volatility estimates shown in Figure 1, LEHD-ACS earnings volatility shown in Figure 2 is lower, although the trend is almost identical. The reduced volatility of the LEHD-ACS series relative to the LEHD series highlights the lower earnings volatility of ACS respondents in general. However, both the LEHD and the LEHD-ACS earnings volatility series are likely biased upwards by workers who move in and out of LEHD-covered employment sectors. The LEHD-ACS UI Covered Employment series shows the impact of removing these workers from the analysis sample, reducing the level of earnings volatility by about 6%, but the trend in earnings volatility is almost identical.

---

[8] Approximately 3% of LEHD year-pair earnings observations in the LEHD-ACS matched sample are likely to have at least some earnings in an employment sector not covered in the LEHD data. See Appendix Table F1 for more details.

The PSID has been the workhorse data set in the earnings volatility literature. A more in-depth comparison of the impact of differences in cross-sectional earnings distributions in the PSID and the LEHD on estimated volatility trends is therefore warranted. Given the large differences between the PSID and LEHD data in the share of workers at the bottom of the earnings distribution, the estimated level of volatility will differ between the two data sources. However, perhaps more important, different trends in the share of workers at the bottom of the earnings distribution and different trends in volatility for these workers have the potential to noticeably affect the trend in overall earnings volatility.

For the set of LEHD workers with earnings in both years of the year-pair, we create two new earnings trims. Starting with the LEHD data trimmed at the $1^{st}$ and $99^{th}$ by-year percentiles of the LEHD earnings distribution, we then also trim LEHD initial and subsequent-year earnings at the PSID by-year $1^{st}$ and $99^{th}$ percentile annual real earnings values. Next, we trim LEHD average annual real earnings for the year-pair at the PSID $1^{st}$ and $99^{th}$ percentile year-pair average real earnings values. Using the PSID ventiles, we also create a set of weights designed such that the weighted LEHD annual earnings distribution is approximately equal to the corresponding PSID average annual earnings distribution.

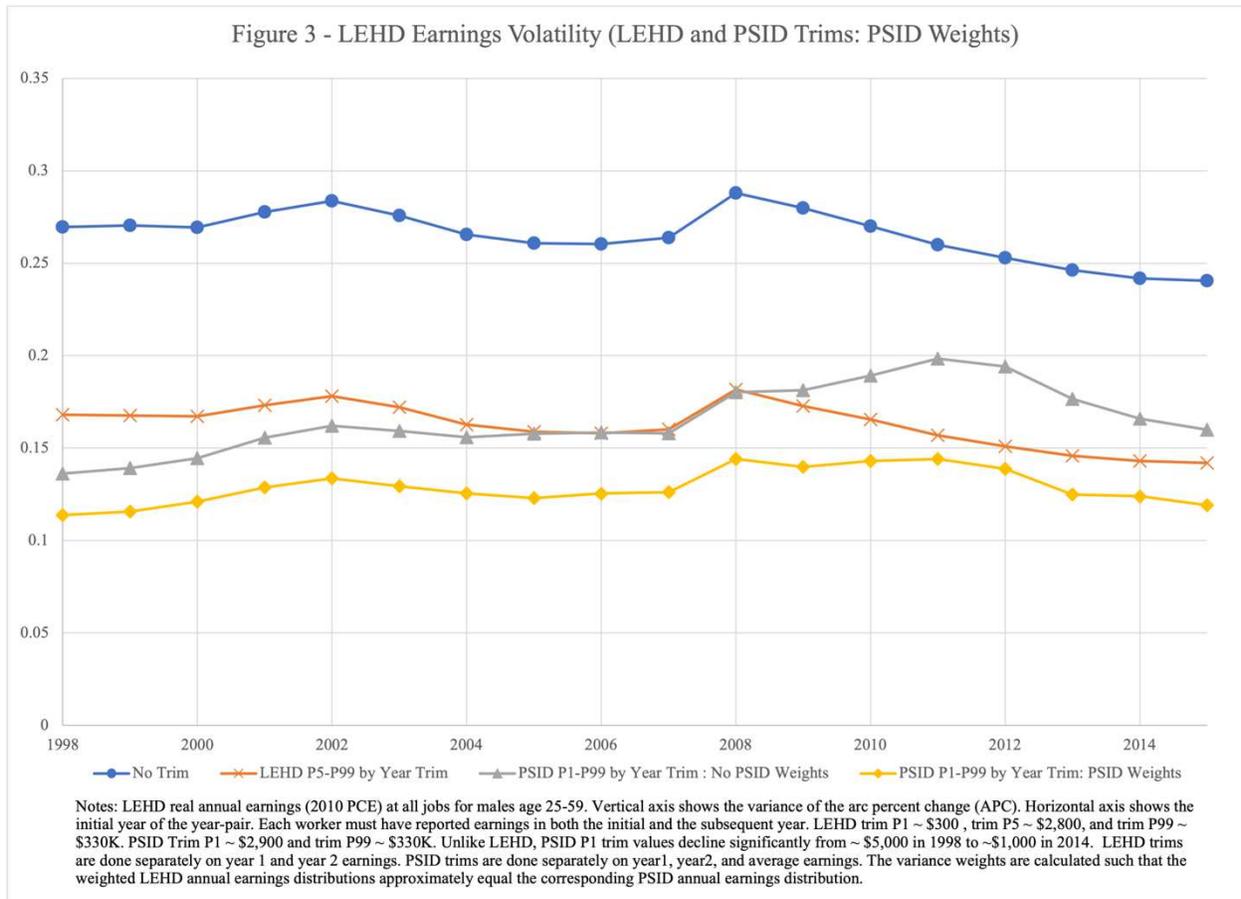

Figure 3 shows the results of applying the various combinations of PSID trims and PSID weights to the LEHD data. All the series shown in Figure 3 use LEHD earnings data, the PSID data are only used to determine weights and trim values. As a reference point, we plot the no weight, no trim (label: No Trim) variance of the arc percentage change series shown in Figure 1. Applying just the PSID trim (label: PSID P1-P99 by Year Trim: No PSID Weight), the overall level of earnings volatility falls and the trend in volatility changes from decreasing to increasing. Applying both the PSID trim and the PSID weights (label: PSID P1-P99 by Year Trim : PSID Weights) further reduces the level of volatility relative to the PSID-trim-only series, while also moderating the increasing trend. For comparison, we also show an additional series without PSID trims and weights where the initial and subsequent earnings values are trimmed at the LEHD 5th and 99th percentile by-year earnings values (label: LEHD P5-P99 by Year Trim). Both series show the impact of removing low earnings values, although the trend changes only when using the PSID trims and weights. Unlike in the LEHD data and other trimming strategies where the by-year trim values are relatively stable over time, the 1st percentile by-year trim values for the PSID decrease significantly over the analysis period and it is this feature of the PSID earnings distribution that generates the difference.[9]

### c. Volatility by Initial-Year Earnings

Above we have focused exclusively on estimating national earnings volatility trends, however our large sample size allows us to estimate separate results for subpopulations. In our final two figures we show results separately for workers at different levels of the initial-year earnings distribution. Workers in the bottom 25% of the initial-year earnings distribution are responsible for approximately 65% of earnings volatility and may have trends in volatility that differ substantially from workers at the middle and top of the initial-year earnings distribution. In Figure 4 we show LEHD earnings volatility estimates within initial-year earnings bin. Earnings volatility levels follow a U-shaped pattern as we move up the earnings distribution; earnings volatility is highest at the bottom, decreasing as earnings increase except at the very top where volatility increases. However, perhaps more importantly, earnings volatility trends are decreasing within each earnings bin. This result makes it extremely unlikely that any type of earnings bin reweighting would produce a positive earnings volatility trend.

---

[9] See Appendix G for additional details, including graphs of the trim values and their impact on the sample of LEHD workers.

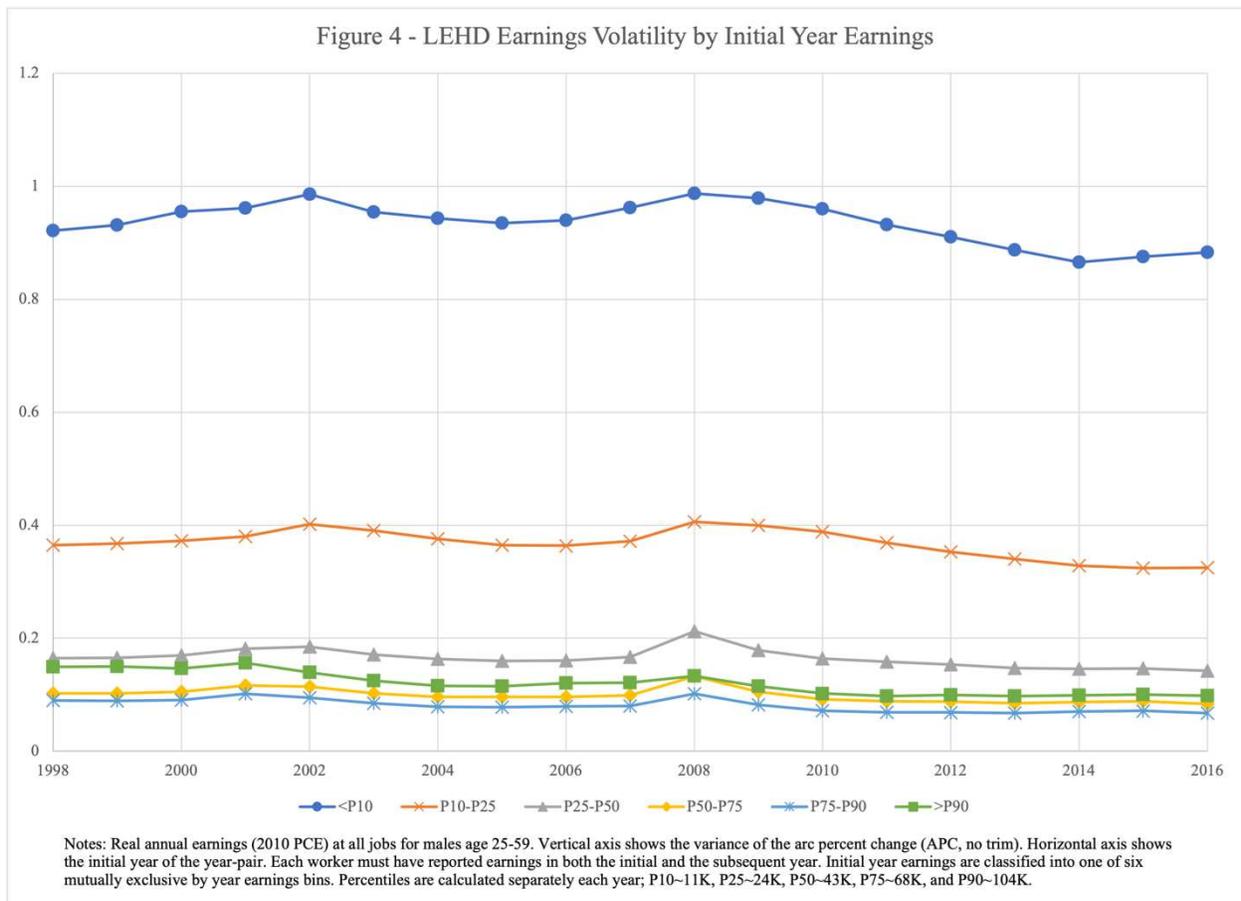

Although it is hard to tell from Figure 4, there are interesting differences across earnings bins in the relative (compared with 1998) changes in earnings volatility. For example, in Figure 5 we see that the largest relative reduction in earnings volatility occurs at the top of the earnings distribution.

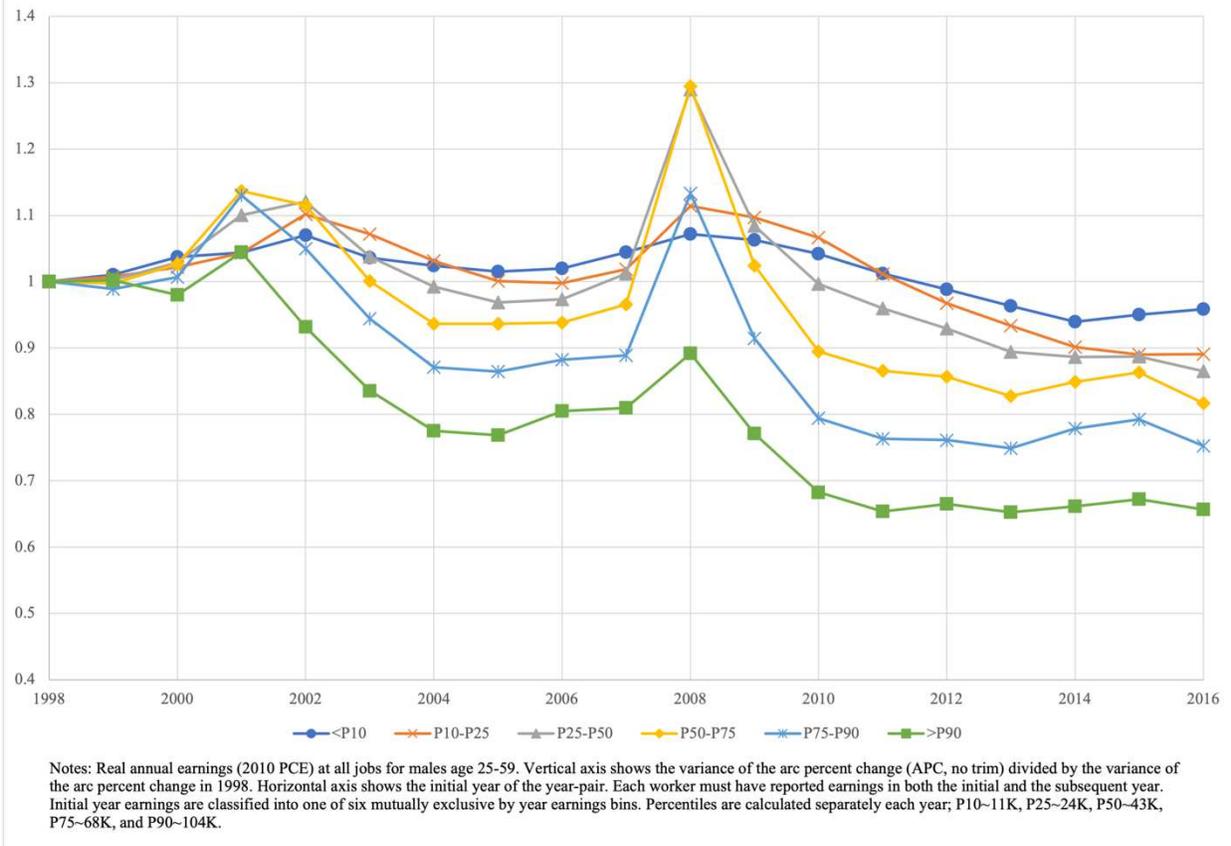

Earnings volatility at the top of the earnings distribution in 2016 is approximately 35% lower than it was in 1998. In contrast to Figure 4, there is an almost monotonic decrease in the relative decline in inequality as we move down the initial-year earnings distribution. Although workers at the bottom of the initial-year earnings distribution have seen a large absolute decrease in earnings volatility, the relative decline is small compared with workers higher up the initial-year earnings distribution who have low absolute levels of volatility yet also have the largest relative reductions in volatility.

## IV. Conclusions

Our estimates of LEHD earnings volatility trends are robust. The large sample sizes combined with our efforts to create a consistent dynamic frame result in an accurate representation of the bottom tail of the earnings distribution, a requirement when estimating earnings volatility trends. Although the relatively large number of low earning workers in LEHD data generate relatively high levels of earnings volatility, a simple by-year 5$^{th}$ percentile trim reduces our earnings volatility estimates to levels roughly comparable to various survey data sources such as the PSID, CPS, and SIPP.

At the national level, earnings volatility trends show a strong decline after the Great Recession. However, except for workers at the bottom of the initial-year earnings distribution (<25$^{th}$ percentile), the downward trend in volatility likely started earlier, after the previous recession in 2001. The downward trend in earnings volatility after 2001 is particularly strong for workers at the top of the earnings distribution.

Although volatility is declining overall, the decline is occurring most slowly for workers least able to self-insure. Future work would do well to exploit the dense nature of administrative data sources to better understand earnings volatility dynamics, keeping in mind that some level of earnings volatility is desirable.

# Online Appendices

**Appendix A – Data**

Table A1 shows information about the available data for each of the fifty states, plus the District of Columbia.[1] By 1995Q1 25 states, including the larger states with relatively high shares of high-earning workers (Illinois, California, Florida, New York, and Texas) are available in the LEHD infrastructure file system, representing about 68% of 2012 BLS Quarterly Census of Employment and Wages (QCEW) Month 1 employment. Between 1995Q2 and 1998Q1 fourteen new states enter the LEHD infrastructure file system, representing an additional 21% of 2012 BLS QCEW Month 1 employment. Thus, by 1998Q1 the LEHD infrastructure consists of 39 states representing approximately 89% of 2012 BLS QCEW month 1 employment. Given the relatively large share of employment entering the LEHD infrastructure (21%) between 1995Q1 and 1998Q1 and the sensitivity of the variance of the change in earnings to the composition of the left tail of the earnings distribution we have chosen to take a conservative approach and start our analysis in 1998Q1.

State entry continues over the ensuing years until 2003Q3 when the final state, Mississippi, is available for analysis. The complete data for all fifty-one states are available through 2015Q4, however in 2016Q1 Missouri exits the LEHD infrastructure file system. Alaska and South Dakota also exit prior to the end of the sample period in 2016Q3 and 2017Q2, respectively. To capture the different epochs of data availability, we create three sample regimes. Each sample regime consists of a set of states available each quarter over a pre-defined time period. The three sets of states and the time periods for each regime are chosen such that state entry/exit are forced to occur at two discrete points in time, thus allowing for a clear presentation of the cumulative effect of state entry/exit. The three sample regimes are defined in Table A1. The first sample regime consists of 39 states with data available beginning no later than 1998Q1 and ending no earlier than 2015Q4. The second sample regime consists of all fifty-one states beginning in 2004Q1 and ending in 2015Q4. The third sample regime consists of 48 states beginning in 2004Q1 and ending in 2017Q4.

When studying earnings volatility, we are primarily concerned with calendar year to calendar year changes in earnings. To facilitate this, we transform the annual earnings data (1998-2017) into a year-pair dataset (1998-2016) where each observation contains information from the initial and the subsequent year. Table A2 shows counts (rounded to four significant digits) of our analysis sample of active, prime-age, male, eligible-workers by year-pair, broken down by labor market status (active both years or active in only one year) and sample regime. For all results except Figure A2, we use a composite sample constructed from the three sample regimes. The composite sample is constructed each year by choosing the available sample regime with the largest number of states. Applying this rule results in a composite sample consisting of regime 1 from 1998-1999 to 2003-2004, regime 2 from 2004-2005 to 2014-2015, and regime 3 from 2015-2016 to 2016-2017.

---

[1] In the discussion that follows we treat D.C. as a state.

We use the sample regimes to assess the impact of two alternating periods of state level missing data. At the beginning of our analysis period, twelve states representing approximately 11% of QCEW 2012 month 1 employment are missing, followed by a complete data period, while near the end of our sample three states that make up about 2.5% of employment are missing. If states are missing completely at random, then the level and trend of earnings volatility should not noticeably be affected during periods where states are missing relative to periods where the data are complete. In Figure A1, we show two different measures of earnings volatility: the first is the difference in log earnings (DLE, no trim), $l_{it}$, and the second is the arc percentage change (APC, no trim), $a_{it}$. For both measures, the calculation includes only workers with earnings in both the initial and the subsequent year. Three separate series are shown, one for each sample regime. The sample regime series overlap during the period 2004 to 2014, allowing for an assessment of the effect of state entry/exit. During the overlap period all series have identical trends and any difference in levels appears to be directly related to the number and size of states omitted from each regime. Given the small share of missing employment it is not surprising Regime 3 overlays the complete data Regime 2.

Comparing Regime 1 with Regime 2, we observe a slight difference in volatility levels for the Regime 1 DLE volatility measure relative to the Regime 2 DLE volatility measure from 2004 to 2009, although in practice the difference is unlikely to significantly affect inferences. The differences across regimes for the APC measure are even smaller, suggesting much of the difference in volatility across regimes is due to large changes in earnings, which typically occur for workers with very small earnings in one year of the year-pair. For all practical purposes earnings volatility estimated using a composite sample composed of the largest available regime each year is virtually identical to results estimated separately for each regime.

Estimated annual real earnings distributions differ depending on the specific data sources used, which can have a noticeable impact on estimated earnings volatility, especially if the differences are concentrated in the lower (left) tail of the earnings distribution. Our primary analysis dataset is based on state Unemployment Insurance (UI) records. Compared with survey data sources, UI data contain a relatively large number of low-earning workers. In this section, we document the differences in the earnings distribution between the LEHD, PSID, and a sample of LEHD earnings for workers who appear in the American Community Survey (the LEHD-ACS sample).

LEHD earnings data are reported directly by the employer (often electronically) and have broad coverage of both the private sector and state/local government. However, unlike household survey data sources LEHD UI data do not include jobs for federal and self-employed in not incorporated firms.[2] Although household survey data covers a potentially broader range of employment arrangements, earnings reports from household surveys take a less direct route to the analyst. To receive an accurate response the frame of households must cover the target population. If the household is selected, the respondent must both answer the survey and correctly answer detailed demographic and historical earnings questions for all in-sample persons in the household.[3] If there is a breakdown anywhere in the chain, the survey earnings reports may either be missing or inaccurate. For stable workers in a single-earner household, the respondent is both more likely to be the actual worker and have similar earnings each year, making it easier to recall previous-year annual earnings. However, in households with multiple workers and/or less stable employment histories there are many more challenges. First, the respondent is less likely to account for the majority of earnings in the household, potentially forcing that person to estimate earnings for other household members. Second, if the respondents have multiple employers with varying earnings from year-to-year it may be difficult to accurately recall previous year earnings. Third, even if the respondent has accurate earnings information for all jobs and for all in-scope persons, the survey may be designed to record earnings only from the primary or dominant job, resulting in an incomplete picture of total annual earnings.[4]

---

[2] Although some federal worker earnings information is available at LEHD, the time period and coverage varies across the entire analysis period. See AMZ for a more detailed explanation of why we excluded federal workers from this study.

[3] Both the ACS and the PSID ask one respondent per household to complete the data for all in-scope persons, although most researchers using the PSID typically only analyze earnings responses for the actual respondent or head of household.

To better understand the magnitude of the differences in the earnings distributions across surveys we construct two additional samples, the first from the PSID and the second combining the workers sampled into the ACS with their LEHD earnings data. We use initial-year real annual earnings for workers with earnings in both the initial and the current year, pooling the years from 2002-2014. The PSID sample was constructed using the online PSID tool and is designed to match as closely as possible both the LEHD UI data sample (males, age 25-59) and the sample selection rules used in Moffitt and Zhang (2020). The PSID is a biennial survey, giving us potentially seven years of earnings for each respondent. The resulting analysis sample size is approximately 15,200 earnings observations. We constructed the LEHD-ACS sample by matching every person from the 2002 to 2014 ACS with that person's year-pair work history from the LEHD analysis sample used in this paper. The resulting LEHD-ACS analysis sample contains about 622 million worker year-pair earnings observations compared with 8.7 billion such observations in the composite LEHD sample.

We estimate the earnings distribution by calculating histograms for each data source. Each observation is placed into one of twenty bins with boundaries approximately equal to the ventiles of the 2002-2014 LEHD composite sample. Figure A2 shows the resulting earnings distributions for the three data sources. Relative to the LEHD and LEHD-ACS samples, the PSID earnings distribution is shifted to the right (except at the very top). The differences are especially pronounced for workers with earnings below $25,000. The LEHD data has about 26% of workers with earnings less than $25,000 and the LEHD-ACS sample has about 24%, while the PSID has only 14%. The large number of low earning workers in the LEHD data will result in higher earnings volatility estimates ceteris paribus than estimates using household data sources.

**Appendix B – Age Distribution**

Ignoring increases in volatility during recessions, earnings volatility declined from 1998 to 2016. One plausible explanation for the decrease in volatility is a shift in the age distribution to the right and/or increased labor force participation among older workers. If older workers have higher earnings and more stable employment this would generally imply a decrease in earnings volatility. Figure B1 shows the age distribution for active workers by year. The age distribution below the median is mostly stable over time, however there is a mild shift to the right in the age distribution above the median. For example, in 1998 the 75$^{th}$ percentile has an age of 47 compared with 49 in 2016.

---

[4] Obtaining accurate survey-based earnings responses can be challenging, although if the issues mentioned above affect data collection consistently over time then earnings volatility trends would be relatively unaffected. Consistency is the key when estimating trends, for example even procedures that improve data collection may affect the trend if they change the composition of the sample.

Although a shift in the age distribution is suggestive, for our purpose we need to determine if a shift in the age distribution and/or the age-earnings profile has a significant effect on earnings variability. To test this hypothesis, we estimate a model of earnings each year, including an age quadratic, using the estimated age-adjusted residuals $\hat{\varepsilon}_{it}$ to calculate the difference in the age-adjusted log real annual earnings residuals $r_{it} = \hat{\varepsilon}_{it+1} - \hat{\varepsilon}_{it}$. In Figure B2, we clearly see the variance of the difference in the age-adjusted residuals is virtually identical to the variance of the difference in log earnings.[5]

**Appendix C – Non-Parametric Earnings Volatility**

Our main results exclusively use the variance as a measure of dispersion. The variance is sensitive to outliers, which we address using the arc percentage change approach and various trimming strategies. An alternative to using the variance is to instead use a non-parametric approach such as the difference between the 90$^{th}$ percentile and the 10$^{th}$ percentile of the difference in log earnings (DLE) distribution. In Figure C1 we present estimates using both measures of earnings volatility. The levels differ (as expected), however the trends in the two series are almost identical. For completeness, we also present various percentiles of the DLE distribution in Figure C2.

**Appendix D – Trim Values and Selected Trimming Approaches**

For the results presented in the body of the paper we use percentiles of the LEHD earnings distribution to trim or remove year-pair observations with either low and/or high annual earnings. The left tail trimming (low earnings observations) has a large impact on the estimated level of volatility, especially when using the difference in log earnings (DLE) measure, although the trends for both the DLE and APC measures are generally not sensitive to the choice of trimming strategy. In Figure D1, we show the time series for various left tail trimming strategies. The LEHD P1 and P5 trimming values are calculated using LEHD data, while the other trimming strategies use external data sources such as the federal minimum wage to set the minimum level of annual earnings. Each of the left tail trim series is flat (LEHD P1, Kopczuk et al. (2010), Guvenen et al. (2014) and Bloom et al. (2017)) or has an upward trend (LEHD P5, Sabelhaus and Song (2009,2010)) in real earnings. In Figure D2 we show APC estimates for each of the trimming strategies. All of the series show a clear downward trend, although the estimated level of earnings volatility and the intensity of the downward trend varies somewhat over the series. The trends for the "flat" trims (LEHD P1, Kopczuk, Guvenen and Bloom) show the smallest decrease in earnings volatility over the analysis period, while the trends for the "upward" trims (LEHD P5, Sabelhaus and Song) show the largest decrease over the analysis period.

---

[5] We also estimated a more general form of the age adjustment directly on the difference in log earnings with similar results.

**Appendix E – Earnings Volatility Including Exiters and Entrants**

In the paper we focus exclusively on workers with reported earnings in two consecutive years, however many workers are inactive for relatively long periods of time, especially during recessions. Our annual earnings measure is based on calendar year earnings; therefore, there is not a direct relationship between observed annual earnings and long inactivity spells. Depending on the start and end dates of the worker's job, a worker with a similar gap in employment may have earnings in two adjacent years while another worker may not. Although we cannot resolve this timing issue, we can increase our capture of long duration periods of inactivity by allowing for year-pair observations where a worker has positive reported earnings in only one of the two years.

We label workers with reported earnings only in the initial year as "exiters," and workers with reported earnings only in the subsequent year as "entrants." Workers with reported earnings in both years are labeled as "stayers." In a typical year (composite sample average) approximately 89% of workers are stayers, 6% of workers are exiters, and 5% of workers are entrants. Graphs of the shares of exiters, entrants, and stayers by year relative to the sample averages are shown in Figure E1.

The recessions in 2001 and 2008 were both associated with large changes in the share of exiters and entrants. During recessions, exiters increase and entrants decline, and there are small changes in the share of stayers. During the 2001 recession the elevated share of exiters and reduced share of entrants begins in 2000 and ends in 2003. During the 2008 recession the divergence begins in 2006 and ends in 2009, although both exiters and entrants are noticeably elevated in 2009 relative to 2003. The average share of exiters being greater than the average share of entrants implies the flow of exiters is greater than the flow of entrants and that net exiters (exiters – entrants) is positive. The dip in the share of stayers is also noticeable during the great recession, especially given their large share. Although entrants are elevated from 2009 to 2013 and exiters are depressed from 2011 forward, employment does not fully recover from the great recession until 2014. Finally, we also observe less mobility into and out of active status from 2014 forward with more stayers, fewer exiters, and fewer entrants.

In Figure E2 we show estimates of the variance of the arc percentage change—one including entrants and exiters, and the other excluding them. The results highlight the significant share of earnings volatility missed by not including exiters and entrants. Including workers with earnings in only one year of the year-pair substantially increases estimated earnings volatility—from about 0.27 to 0.70 in year 2000, for example. As predicted from our discussion of Figure E1, the peaks are larger during recessions as entrants and exiters increase, while stayers decrease.

**Appendix F – Earnings Volatility by Class of Work**

LEHD UI data does not include federal jobs or self-employment at a not incorporated firm. To better understand the impact of these "missing" jobs on our LEHD volatility estimates we match ACS respondents with LEHD earnings responses by worker year. Unlike the first matched ACS sample used to generate Figure A2, we match by person year to retrieve contemporaneous information about a worker's class of work. Specifically, we use the ACS "class of work" responses to identify LEHD workers employed in the types of jobs not covered by LEHD UI data.

Table F1 contains estimates from the public use ACS and the LEHD-ACS matched sample. Table F1 has two panels, the first panel includes LEHD year-pair observations with positive LEHD earnings in at least one year, while the second panel includes only observations with positive LEHD earnings in both years. The sample size for the first panel is 728 million worker year-pair observations and the sample size for the second panel is 677 million worker year-pair observations. The sample used for the first panel includes workers that completely enter/exit the UI covered population in one year of the year-pair, while the second panel requires workers to have at least some UI covered earnings in both years. The first panel sample is useful to better understand total mobility between the UI covered and not covered sectors, although our main volatility results require UI covered earnings in both years.

The worker share columns show the proportion of workers in both the public ACS and the matched LEHD-ACS sample by class of work. In the public ACS data about 10% of workers are employed in the federal government or are self-employed in a not incorporated firm (the bottom two rows in each panel). In the LEHD-ACS matched sample 4.5% (top panel) to 3.1% (bottom panel) of the workers are likely employed at some point during a two-year period in both a UI-covered and a UI not covered sector. Although most UI not covered workers (~55%) do not also work in the UI covered sector during a two-year period, a substantial number transition between or are simultaneously employed both in a UI covered and not-covered sector. LEHD UI data will under-report earnings for these workers in one or both years of the year-pair. The median annual earnings columns show the ACS reported annual earnings and the LEHD annual earnings for workers in the matched LEHD-ACS sample. Median annual earnings reported on the ACS and the LEHD data are similar for private sector, state, and local workers, sectors we expect to be well covered by LEHD data. However, large discrepancies in annual earnings exist for self-employed and federal government workers, results consistent with our expectations.

The missing earnings from UI not covered jobs has large impacts on estimated volatility for workers employed in both the UI not covered and UI covered sectors. Earnings volatility estimates for workers with UI not covered jobs are two to five times higher than private sector earnings volatility estimates for workers with zero earnings in one year of the year-pair. However, the second panel results, which are more relevant for our volatility estimates, show noticeably less impact than we would expect based on the first panel results alone. Both the share of workers in the not covered sectors and earnings volatility are lower. Overall, the expected impact on earnings volatility is relatively small. See Figure 2 for earnings volatility trend estimates excluding year-pairs with suspected UI not covered employment.

**Appendix G – Earnings Volatility by Initial-Year Earnings Bin**

In section III.b. we conduct an experiment designed to adjust our earnings volatility measures for differences in the real annual earnings distribution between the PSID and the LEHD data. The experiment is designed as follows; in the first step, we select a sample of LEHD records comparable to what might be expected to appear on the PSID by applying PSID trimming rules to the LEHD data. In the second step, we use the set of selected records to estimate a weighted earnings volatility measure, where the weights are calculated such that the weighted LEHD earnings distribution matches the unweighted PSID earnings distribution.[6]

Given the large difference between the left tail of the earnings distribution in the PSID and the left tail of the LEHD data (see Figure A2) and the dominant share of earnings volatility contributed by low earning workers it is worthwhile explaining in detail exactly how the left tail trimming was implemented. The trimming is implemented in multiple stages, with each trimmed value used as an input to the next layer of trimming. As a first step, LEHD real annual earnings are trimmed using the LEHD by-year 1st and 99th percentiles. Next, the initial-year and subsequent-year LEHD real annual earnings are trimmed using the by-year PSID 1st and 99th initial and subsequent-year percentiles. In the final trim, the average of initial and subsequent-year LEHD earnings are trimmed at the min/max values of the PSID average real annual earnings distribution (by year). The trims are designed to generate an LEHD sample with the same range of earnings as used in the PSID.

The weights are constructed using PSID year-pair average real annual earnings ventiles. The weights reflect the relative difference in the proportion of records for each data source in each ventile based bin. For bins where the LEHD data has a relatively large share of earning records, the weight is less than one and is greater than one for bins where the PSID has a relatively large share of earnings records. The earnings distribution comparison in Figure A2 provides a rough picture of the weights. The weights are typically less than one below around $30,000 and greater than one above that amount (except for the extreme right tail where the weights are about one).

The first layer of trimming using LEHD based trimming values removes approximately the as designed 2% of real annual earnings records each year, while the second and third layers of trimming have a more substantial effect that changes over time. The second layer of trimming has a significant impact due to the increasing over time number of LEHD workers with average real annual earnings above the PSID left tail earnings trim values. In Figure G1 we show the PSID P1 values compared with the approximate 1st and 5th percentile LEHD left tail trim values. What stands out immediately is the strong downward trend in the PSID 1st percentile annual earning trim values from 1998 to 2012 compared with the LEHD percentiles which are relatively stable over the period.

---

[6] The PSID trim values and the ventiles used to construct the weights are from Moffitt and Zhang (2020).

The large difference in the left tail trim values results in a similarly large difference in the percent of the LEHD baseline sample selected each year. The overall effect of applying the combined three layers of PSID based trims on estimated earnings volatility changes the downward trend in earnings inequality found in the baseline sample to an increasing earnings volatility trend (see Figure 3), simply by including a larger fraction of low earning high volatility workers in the sample each year from the beginning of the analysis period to 2012.

Using the same initial-year earnings bins as in Figure 3, in Figure G2 we show within bin earnings volatility estimates after applying the PSID 1$^{st}$ percentile trim. Comparing Figure 3 with Figure G2 we clearly see the impact of applying the PSID trim. Earnings volatility trends for the two lowest earnings bins change from a downward trend in Figure 3 to upward trends in Figure G2. The impact of applying the PSID trims is especially strong for LEHD workers with initial-year earnings in the bottom 10 percent. Figure G3 shows the change relative to 1998 for each within bin earnings inequality series. Once again, the impact is largest for workers at the bottom of the initial-year earnings distribution, although all earnings bins are affected.

When interpreting the within bin earnings volatility estimates it is useful to know whether a weighted average of the within bin variance is a good proxy for the overall variance. Figures G4 and G5 show the results of decomposing overall earnings volatility into two components. The first component is the expected value of the within bin variance series shown in Figure 3 and Figure G2, which is equal to a weighted average of the within bin earnings volatility estimates. The second component is the variance of the mean APC across the initial-year earnings bins. For both Figures G4 and G5, the first component (the weighted sum of the within bin variances) is dominant and shows an almost identical trend to the overall earnings variance, confirming that the first component alone is a good proxy estimate of earnings volatility. The second variance component is relatively flat in Figure 4 but is increasing in Figure G5. The results in Figures G2 and G5 clearly show how the PSID trims substantially changes the earnings volatility results.

Table A1 - State Availability and Sample Regimes

| Count | State | Sample Regime 1 | Sample Regime 2 | Sample Regime 3 | First YYYY:Q Available | Last YYYY:Q Available | Pct 2012.1 QCEW Emp |
|---|---|---|---|---|---|---|---|
| 1 | Maryland | X | X | X | 1985:2 | 2017:4 | 1.83% |
| 2 | Alaska | X | X |  | 1990:1 | 2016:2 | 0.22% |
| 3 | Colorado | X | X | X | 1990:1 | 2017:4 | 1.70% |
| 4 | Idaho | X | X | X | 1990:1 | 2017:4 | 0.45% |
| 5 | Illinois | X | X | X | 1990:1 | 2017:4 | 4.38% |
| 6 | Indiana | X | X | X | 1990:1 | 2017:4 | 2.19% |
| 7 | Kansas | X | X | X | 1990:1 | 2017:4 | 0.98% |
| 8 | Louisiana | X | X | X | 1990:1 | 2017:4 | 1.41% |
| 9 | Missouri | X | X |  | 1990:1 | 2015:4 | 1.99% |
| 10 | Washington | X | X | X | 1990:1 | 2017:4 | 2.12% |
| 11 | Wisconsin | X | X | X | 1990:1 | 2017:4 | 2.08% |
| 12 | North Carolina | X | X | X | 1991:1 | 2017:4 | 2.92% |
| 13 | Oregon | X | X | X | 1991:1 | 2017:4 | 1.23% |
| 14 | Pennsylvania | X | X | X | 1991:1 | 2017:4 | 4.44% |
| 15 | California | X | X | X | 1991:3 | 2017:4 | 11.37% |
| 16 | Arizona | X | X | X | 1992:1 | 2017:4 | 1.85% |
| 17 | Wyoming | X | X | X | 1992:1 | 2017:4 | 0.19% |
| 18 | Florida | X | X | X | 1992:4 | 2017:4 | 5.78% |
| 19 | Montana | X | X | X | 1993:1 | 2017:4 | 0.31% |
| 20 | Georgia | X | X | X | 1994:1 | 2017:4 | 2.90% |
| 21 | South Dakota | X | X |  | 1994:1 | 2017:1 | 0.30% |
| 22 | Minnesota | X | X |  | 1994:3 | 2017:4 | 2.05% |
| 23 | New York | X | X | X | 1995:1 | 2017:4 | 6.49% |
| 24 | Rhode Island | X | X | X | 1995:1 | 2017:4 | 0.35% |
| 25 | Texas | X | X | X | 1995:1 | 2017:4 | 8.10% |
| 26 | New Mexico | X | X | X | 1995:3 | 2017:4 | 0.55% |
| 27 | Hawaii | X | X | X | 1995:4 | 2017:4 | 0.44% |
| 28 | Connecticut | X | X | X | 1996:1 | 2017:4 | 1.26% |
| 29 | Maine | X | X | X | 1996:1 | 2017:4 | 0.43% |
| 30 | New Jersey | X | X | X | 1996:1 | 2017:4 | 2.87% |
| 31 | Kentucky | X | X | X | 1996:4 | 2017:4 | 1.32% |
| 32 | West Virginia | X | X | X | 1997:1 | 2017:4 | 0.52% |
| 33 | Michigan | X | X | X | 1998:1 | 2017:4 | 3.04% |
| 34 | Nevada | X | X | X | 1998:1 | 2017:4 | 0.89% |
| 35 | North Dakota | X | X | X | 1998:1 | 2017:4 | 0.31% |
| 36 | Ohio | X | X | X | 1998:1 | 2017:4 | 3.93% |
| 37 | South Carolina | X | X | X | 1998:1 | 2017:4 | 1.35% |
| 38 | Tennessee | X | X | X | 1998:1 | 2017:4 | 2.03% |
| 39 | Virginia | X | X | X | 1998:1 | 2017:4 | 2.65% |
| 40 | Delaware |  | X | X | 1998:3 | 2017:4 | 0.31% |
| 41 | Iowa |  | X | X | 1998:4 | 2017:4 | 1.12% |
| 42 | Nebraska |  | X | X | 1999:1 | 2017:4 | 0.69% |
| 43 | Utah |  | X | X | 1999:1 | 2017:4 | 0.91% |
| 44 | Oklahoma |  | X | X | 2000:1 | 2017:4 | 1.11% |
| 45 | Vermont |  | X | X | 2000:1 | 2017:4 | 0.22% |
| 46 | Alabama |  | X | X | 2001:1 | 2017:4 | 1.34% |
| 47 | Massachusetts |  | X | X | 2002:1 | 2017:4 | 2.55% |
| 48 | District of Columbia |  | X | X | 2002:2 | 2017:4 | 0.43% |
| 49 | Arkansas |  | X | X | 2002:3 | 2017:4 | 0.86% |
| 50 | New Hampshire |  | X | X | 2003:1 | 2017:4 | 0.47% |
| 51 | Mississippi |  | X | X | 2003:3 | 2017:4 | 0.77% |
|  | Share QCEW 2012.1 | 89.22% | 100.00% | 97.49% |  |  |  |

Notes: Each row represents a state or DC. States are ordered by the quarter their data first became available in the LEHD infrastrucuture files. All states and DC are available by 2003:3. The first three colums show the states in each sample regime (a sample regime is a set of states with the same sample entry and exit dates). Regime 1 excludes states that enter after the start of the analysis period and runs from 1998:1 to 2015:4. Regime 2 is the complete data period and runs from 2004:1 to 2015:4. Regime 3 excludes states that exit before the end of the analysis period and runs from 2004:1 to 2017:4. The last column shows the proportion of each state as a percentage of national 2012 month 1 BLS QCEW employment. The last row shows the share of national 2012 month 1 BLS QCEW employment in each regime.

Table A2 - Sample Sizes by Regime, Active Status, and Year

| Year Pair | Active Both Years | | | | Active One Year | | Total Composite Sample |
|---|---|---|---|---|---|---|---|
| | Sample Regime 1 | Sample Regime 2 | Sample Regime 3 | Composite Sample | Exiters- Active Year 1 | Entrants- Active Year 2 | |
| 1998 - 1999 | 43,180,000 | | | 43,180,000 | 2,874,000 | 2,818,000 | 48,872,000 |
| 1999 - 2000 | 43,830,000 | | | 43,830,000 | 2,926,000 | 2,927,000 | 49,683,000 |
| 2000 - 2001 | 44,350,000 | | | 44,350,000 | 3,168,000 | 2,580,000 | 50,098,000 |
| 2001 - 2002 | 44,040,000 | | | 44,040,000 | 3,574,000 | 2,463,000 | 50,077,000 |
| 2002 - 2003 | 43,620,000 | | | 43,620,000 | 3,516,000 | 2,599,000 | 49,735,000 |
| 2003 - 2004 | 43,390,000 | | | 43,390,000 | 3,074,000 | 2,708,000 | 49,172,000 |
| 2004 - 2005 | 43,560,000 | 48,860,000 | 47,570,000 | 48,860,000 | 3,340,000 | 2,919,000 | 55,119,000 |
| 2005 - 2006 | 43,960,000 | 49,330,000 | 48,030,000 | 49,330,000 | 3,287,000 | 2,936,000 | 55,553,000 |
| 2006 - 2007 | 44,130,000 | 49,540,000 | 48,230,000 | 49,540,000 | 3,287,000 | 2,871,000 | 55,698,000 |
| 2007 - 2008 | 44,130,000 | 49,560,000 | 48,250,000 | 49,560,000 | 3,503,000 | 2,667,000 | 55,730,000 |
| 2008 - 2009 | 43,060,000 | 48,400,000 | 47,120,000 | 48,400,000 | 4,471,000 | 2,228,000 | 55,099,000 |
| 2009 - 2010 | 41,870,000 | 47,110,000 | 45,860,000 | 47,110,000 | 4,060,000 | 3,004,000 | 54,174,000 |
| 2010 - 2011 | 41,960,000 | 47,240,000 | 45,980,000 | 47,240,000 | 3,431,000 | 3,338,000 | 54,009,000 |
| 2011 - 2012 | 42,550,000 | 47,890,000 | 46,630,000 | 47,890,000 | 3,220,000 | 3,206,000 | 54,316,000 |
| 2012 - 2013 | 43,110,000 | 48,510,000 | 47,240,000 | 48,510,000 | 3,116,000 | 3,060,000 | 54,686,000 |
| 2013 - 2014 | 43,680,000 | 49,140,000 | 47,850,000 | 49,140,000 | 2,985,000 | 3,019,000 | 55,144,000 |
| 2014 - 2015 | 44,300,000 | 49,810,000 | 48,510,000 | 49,810,000 | 2,937,000 | 2,944,000 | 55,691,000 |
| 2015 - 2016 | | | 49,110,000 | 49,110,000 | 2,981,000 | 2,809,000 | 54,900,000 |
| 2016 - 2017 | | | 49,630,000 | 49,630,000 | 3,021,000 | 2,855,000 | 55,506,000 |

Notes: An observation is a year pair formed from the initial and the subsequent year. To be included, a worker must have reported earnings in the initial and/or the subsequent year. The worker must also be male, age 25-59 in both years, have no more than 12 jobs during the year, appear on the Census Numident, not be reported dead, and have years in the US>0. Counts are rounded to 4 significant figures. Composite LEHD and LEHD-ACS sample constructed using Regime 1 from 1998-2003, Regime 2 from 2004-2014, and Regime 3 from 2015-2016.
13131313

Table F1 - LEHD Earnings and Volatility by ACS Class of Work

| | | \multicolumn{2}{c}{LEHD Stayers, Entrants, and Exiters (Zero Earnings in One Year)} | | | |
|---|---|---|---|---|---|---|
| | | Worker Share | | Median Annual Earnings | | Volatility |
| ACS Class of Work | LEHD Covered | ACS | LEHD-ACS | ACS | LEHD-ACS | LEHD-ACS |
| Private for-profit | Yes | 0.710 | 0.758 | 43,960 | 43,685 | 0.415 |
| Private not-for-profit | Yes | 0.053 | 0.049 | 46,120 | 44,775 | 0.471 |
| Local government | Yes | 0.055 | 0.070 | 49,310 | 51,405 | 0.230 |
| State government | Yes | 0.036 | 0.040 | 48,720 | 49,470 | 0.277 |
| Self-employed incorporated | Yes | 0.047 | 0.039 | 60,920 | 38,205 | 0.900 |
| Federal government | No | 0.029 | 0.014 | 50,420 | 20,140 | 1.585 |
| Self-employed not incorporated | No | 0.071 | 0.031 | 29,260 | 9,050 | 1.957 |

| | | \multicolumn{2}{c}{LEHD Stayers (Positive Earnings in Both Years)} | | | |
|---|---|---|---|---|---|---|
| | | Worker Share | | Median Annual Earnings | | Volatility |
| ACS Class of Work | LEHD Covered | ACS | LEHD-ACS | ACS | LEHD-ACS | LEHD-ACS |
| Private for-profit | Yes | 0.710 | 0.772 | 45,330 | 45,700 | 0.211 |
| Private not-for-profit | Yes | 0.053 | 0.049 | 47,390 | 47,625 | 0.200 |
| Local government | Yes | 0.055 | 0.073 | 49,830 | 52,255 | 0.122 |
| State government | Yes | 0.036 | 0.042 | 49,370 | 50,660 | 0.126 |
| Self-employed incorporated | Yes | 0.047 | 0.035 | 64,960 | 46,535 | 0.309 |
| Federal government | No | 0.029 | 0.011 | 51,020 | 35,465 | 0.472 |
| Self-employed not incorporated | No | 0.071 | 0.020 | 31,680 | 21,385 | 0.662 |

Notes: The ACS share of workers is calculated using pooled 2010-2016 public use ACS data for males age 16 and over. All other statistics are calculated using a person-year sample of male LEHD workers age 25-59 matched with ACS data from 2001-2016. To be included, a LEHD worker must match to an ACS worker in either year of the LEHD year-pair. Results are calculated using the pooled sample. ACS median earnings are calculated from reported annual wage and salary earnings. LEHD-ACS median annual earnings are calculated using real annual earnings at all jobs (2010 PCE). Volatility is the variance of the arc percent change in

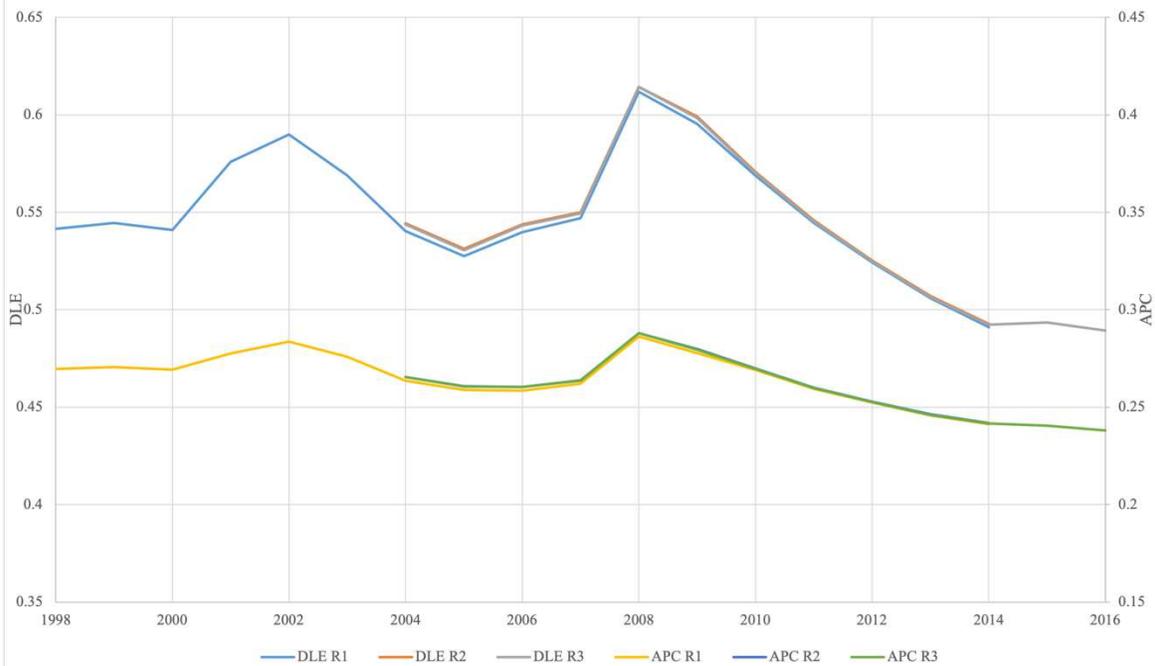

Notes: LEHD real annual earnings (2010 PCE) at all jobs for males age 25-59. The left vertical axis shows the variance of the difference in log earnings (DLE, no trim) and the right axis shows the variance of the arc percent change (APC, no trim). Horizontal axis shows the initial year of the year pair. Each worker must have reported earnings in both the initial and the subsequent year. LEHD data has state entry/exit during the analysis period. Sample regimes are constructed such that sets of states have the same entry/exit dates. Regime 2 (R2) contains all 50 states plus DC. Regime 1 (R1) does not include AL, AR, DC, DE, IA, MA, MS, NE, NH, OK, UT, VT. Regime 3 (R3) does not include AK, MO, and SD.

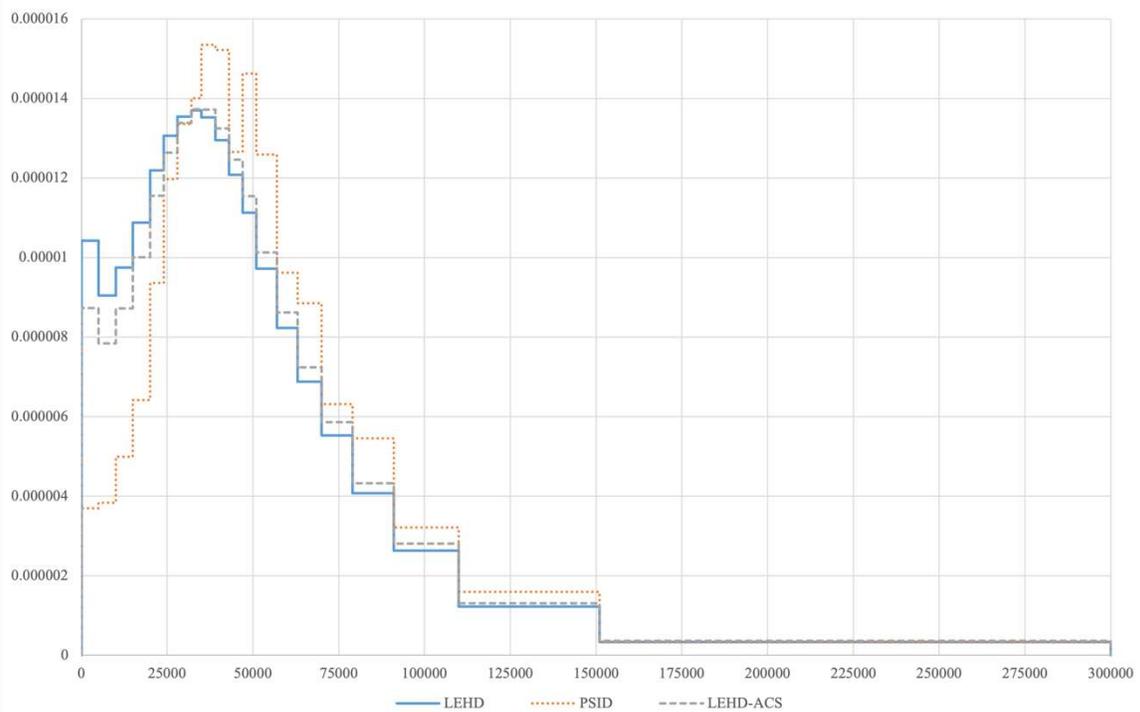

Notes: Real annual earnings (2010 PCE) at all jobs for males age 25-59. Each worker must have reported earnings in both the initial and the next available year of the year-pair (subsequent year for LEHD and LEHD-ACS, initial year+2 for PSID). The LEHD-ACS sample consists of LEHD annual earnings for all in sample LEHD workers that also appear on the ACS during 2002-2014. The PSID sample includes biennial reports of annual wage and salary earnings from 2002-2014.

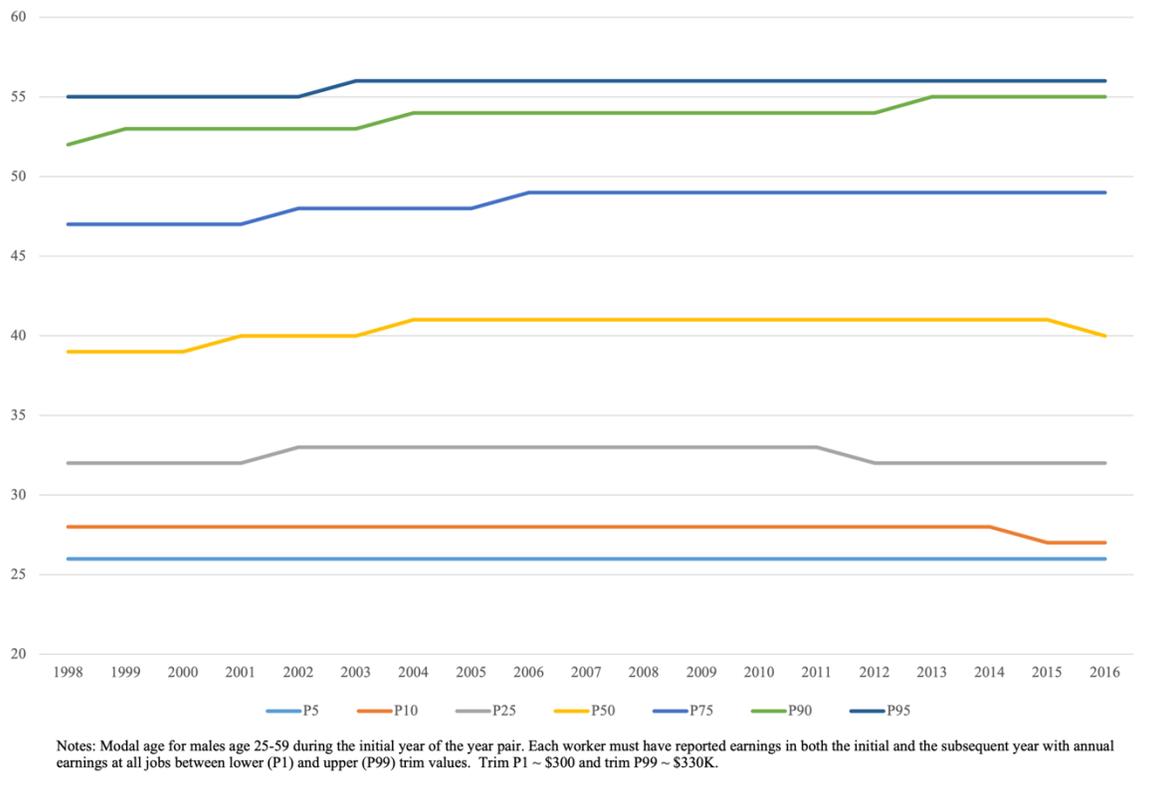

Figure B1 - Age Distribution by Year

Notes: Modal age for males age 25-59 during the initial year of the year pair. Each worker must have reported earnings in both the initial and the subsequent year with annual earnings at all jobs between lower (P1) and upper (P99) trim values. Trim P1 ~ $300 and trim P99 ~ $330K.

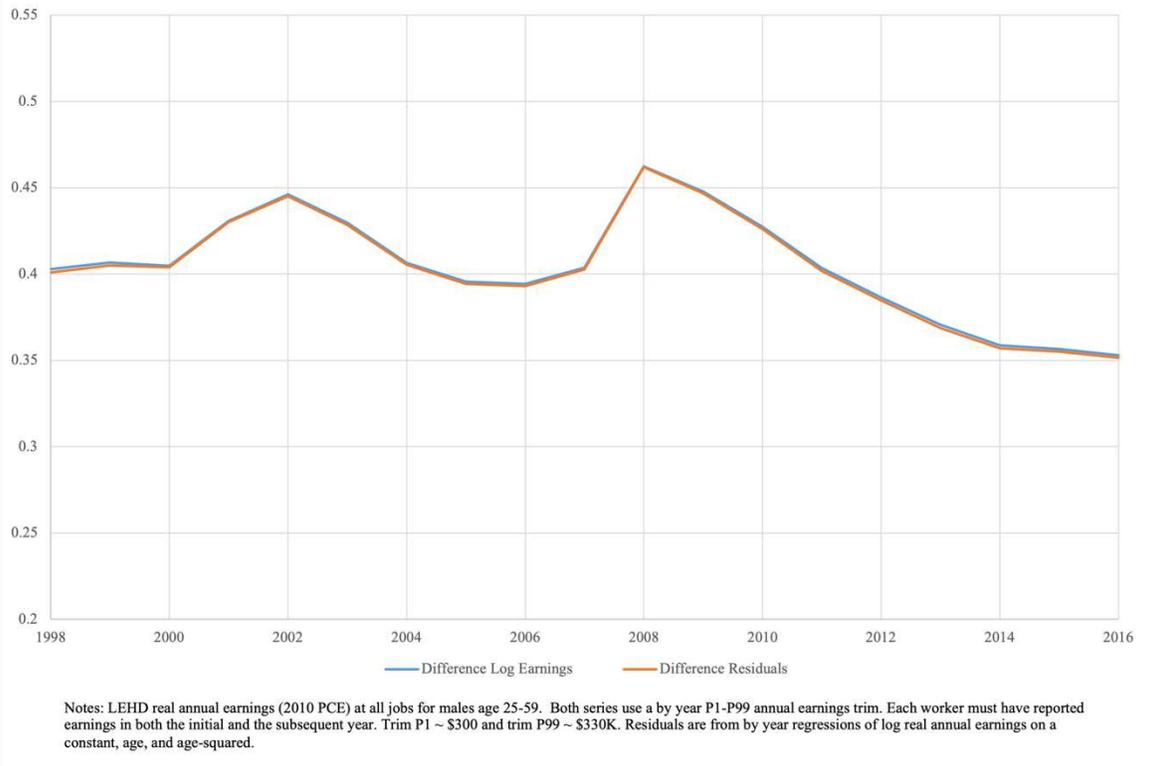

Figure B2 - Variance of the Difference in Log Earnings and the Difference in the Age Adjusted Log Earnings Residuals by Year

Notes: LEHD real annual earnings (2010 PCE) at all jobs for males age 25-59. Both series use a by year P1-P99 annual earnings trim. Each worker must have reported earnings in both the initial and the subsequent year. Trim P1 ~ $300 and trim P99 ~ $330K. Residuals are from by year regressions of log real annual earnings on a constant, age, and age-squared.

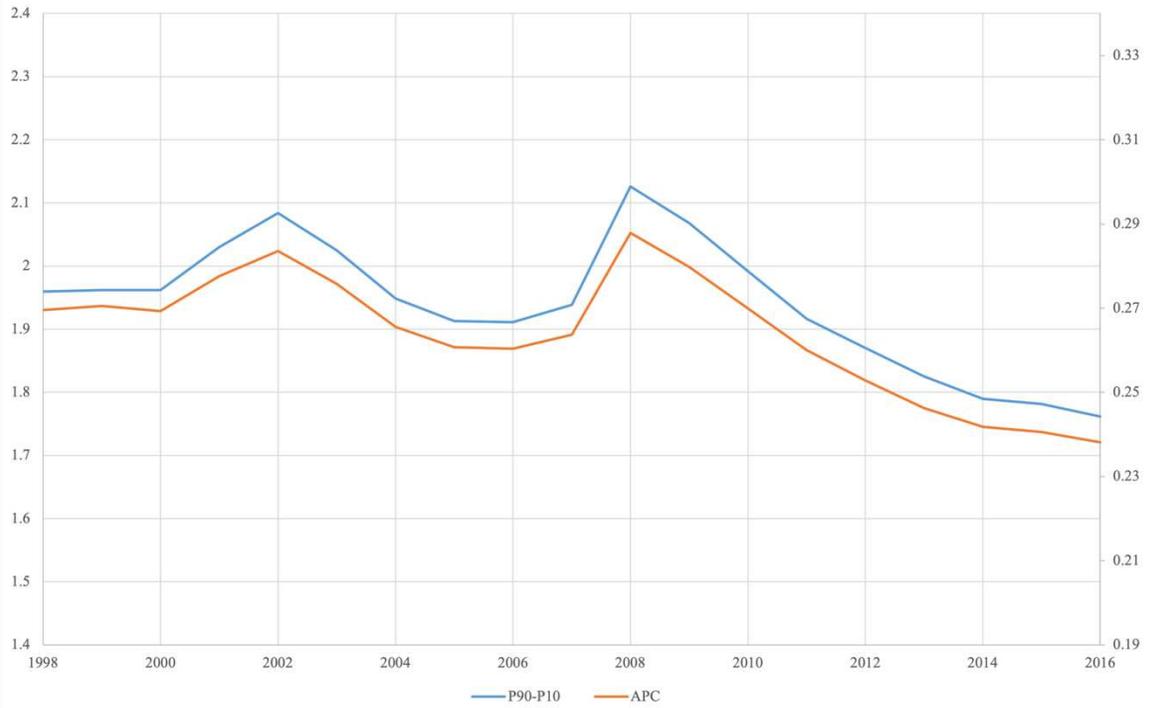

Figure C1 - LEHD Earnings Volatility (P90-P10)

Notes: LEHD real annual earnings (2010 PCE) at all jobs for males age 25-59. Left vertical axis shows the P90 - P10 of the difference in log earnings (DLE, no trim) and the right vertical axis shows the variance of the arc percent change (APC, no trim). Horizontal axis shows the initial year of the year pair. Each worker must have reported earnings in both the initial and the subsequent year

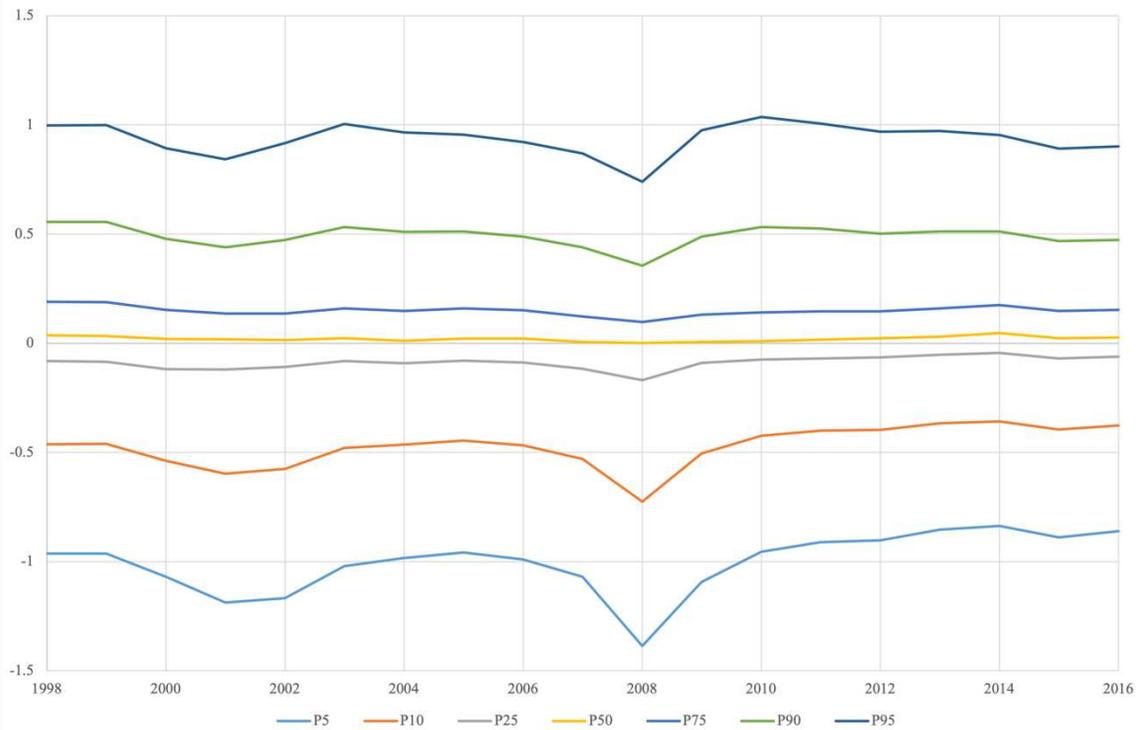

Figure C2 - Percentiles of the Difference in Log Earnings

Notes: LEHD real annual earnings (2010 PCE) at all jobs for males age 25-59. Left vertical axis shows the difference in log earnings (DLE, no trim). Horizontal axis shows the initial year of the year pair. Each worker must have reported earnings in both the initial and the subsequent year

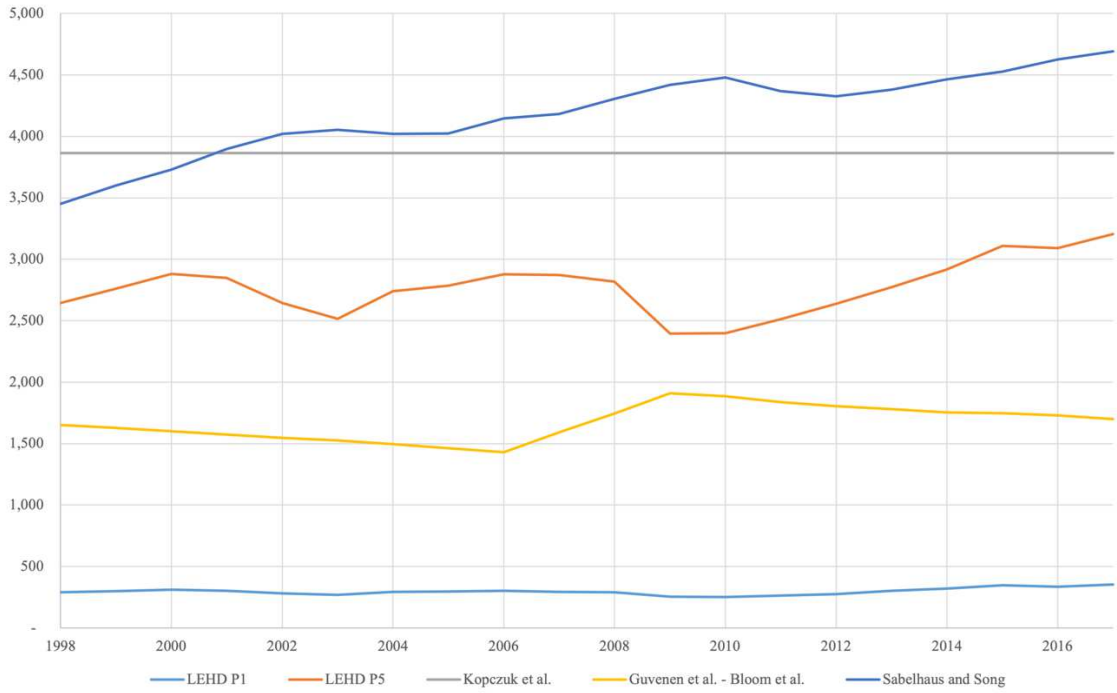

Figure D1 - Various Earnings Distribution Left Tail Trim Values by Year

Notes: Trim values are used to remove low earning/potential high variability workers from the analysis sample. LEHD P1 and P5 calculated using real annual earnings (2010 PCE) at all jobs each year for males age 25-29. Kopczuk et al. method uses a fixed real dollar amount each year (1/4 of full-time, full-year earnings at the 2011 minimum wage, $3,865). Guven and Bloom method uses the real dollar value of working 1/4 of the year, 20 hours a week at the current federal minimum wage. Sabelhaus and Song use the minimum annual earnings each year required to qualify for Social Security credit.

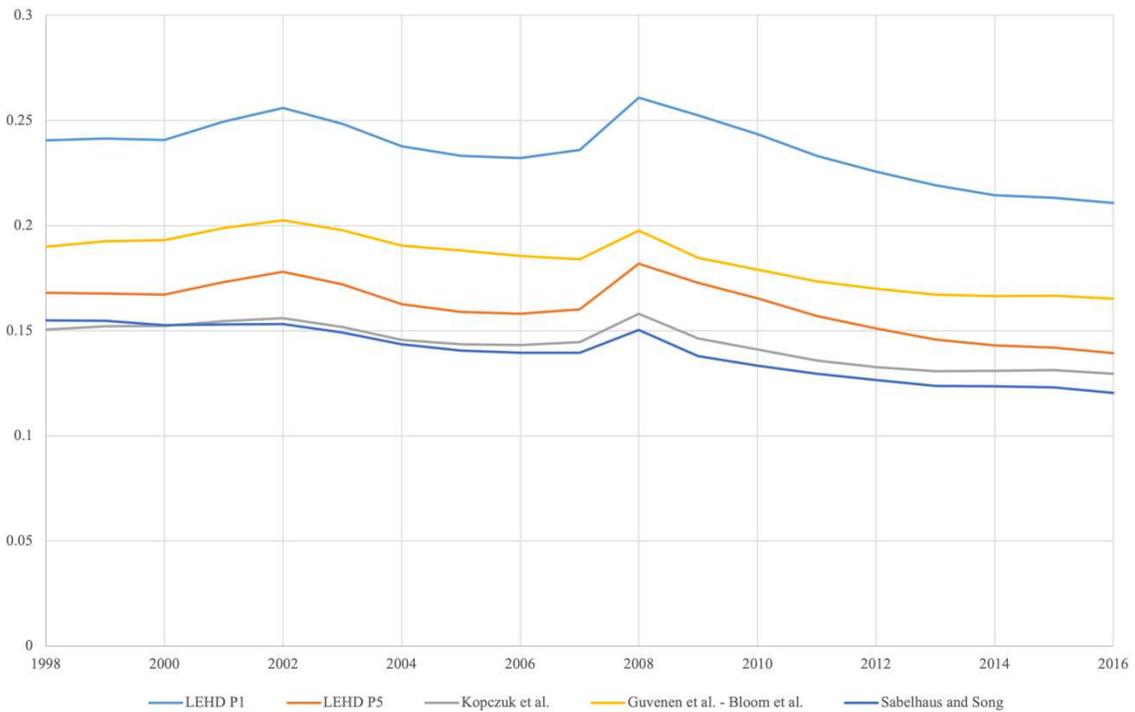

Figure D2 - LEHD Earnings Volatility by Left Tail Trim Type

Notes: LEHD real annual earnings (2010 PCE) at all jobs for males age 25-59. Vertical axis shows the variance of the arc percent change (APC). Horizontal axis shows the initial year of the year pair. All series estimated using LEHD data with the minimum earnings value each year determined by the corresponding trim series shown in Figure D1. The right tail trim values are the LEHD by year P99 ~ 330k for all series.

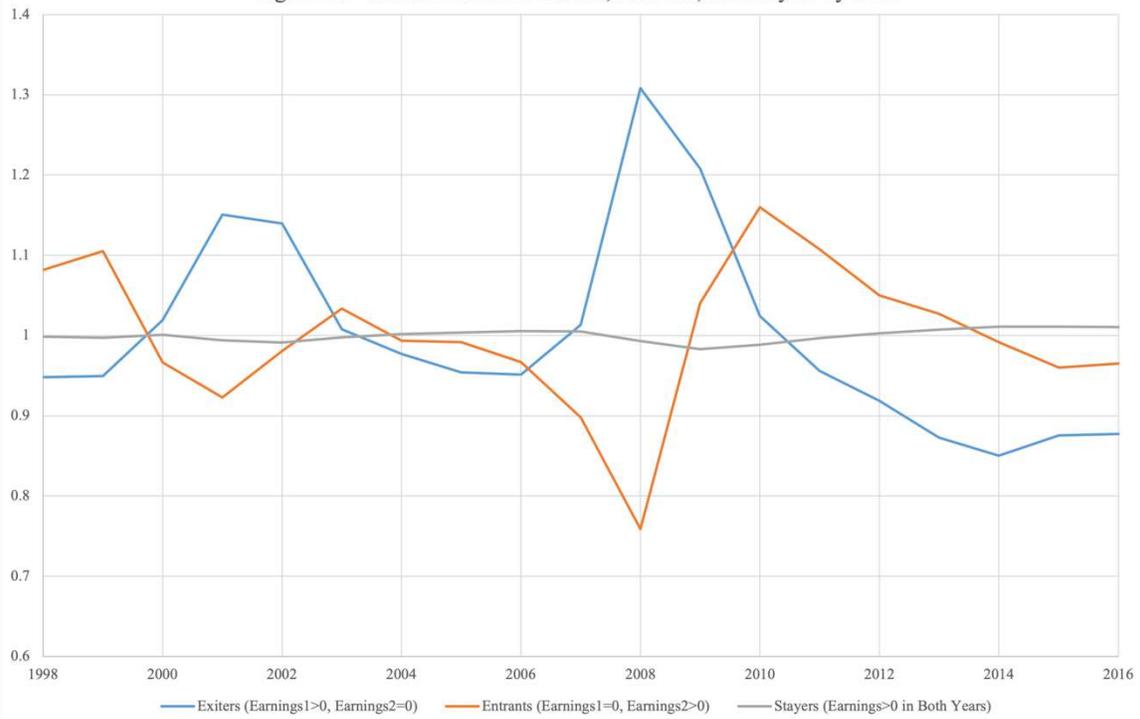

Figure E1 - Relative Share of Exiters, Entrants, and Stayers by Year

Notes: LEHD real annual earnings (2010 PCE) at all jobs for males age 25-59. Each worker must have reported earnings in either the initial and/or the subsequent year. Horizontal axis shows the initial year of the year pair. Shares shown on the vertical axis are relative to the average share across all years. Average share of exiters is 0.062. Average share of entrants is 0.053. Average share of stayers is 0.885.

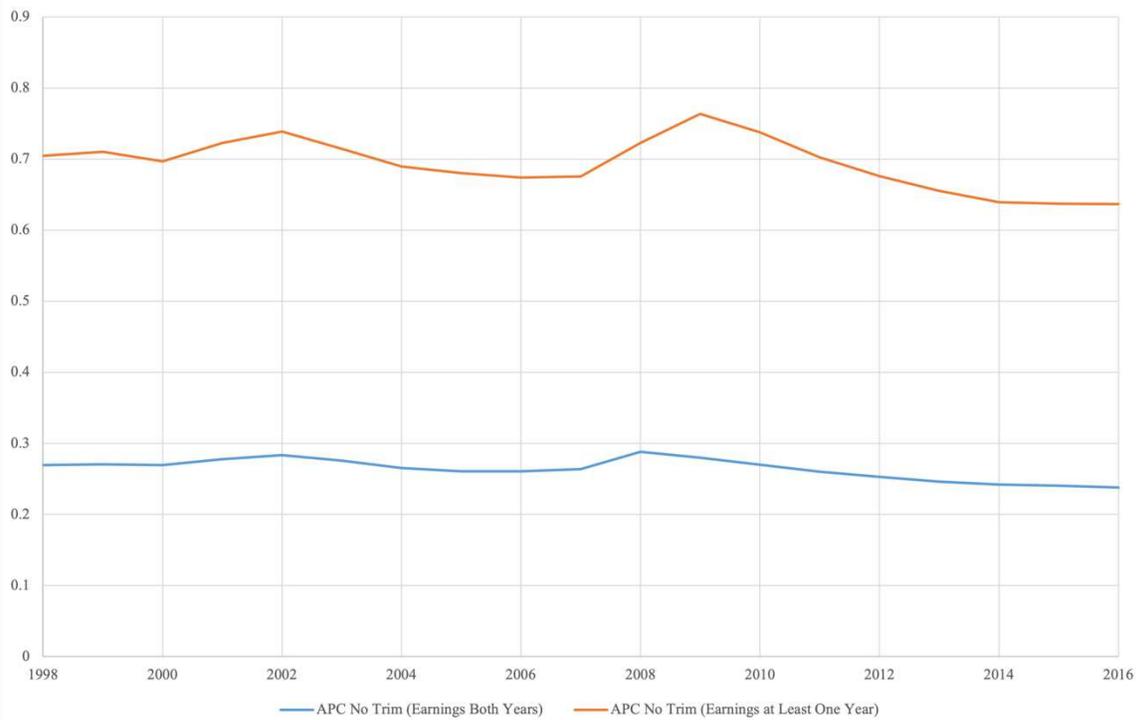

Figure E2 - LEHD Earnings Volatility Including Exiters and Entrants

Notes: LEHD real annual earnings (2010 PCE) at all jobs for males age 25-59. Each worker must have reported earnings in either the initial and/or the subsequent year, zero earnings in one year are allowed. Vertical axis shows the variance of the arc percent change (APC, no trim). Horizontal axis shows the initial year of the year pair.

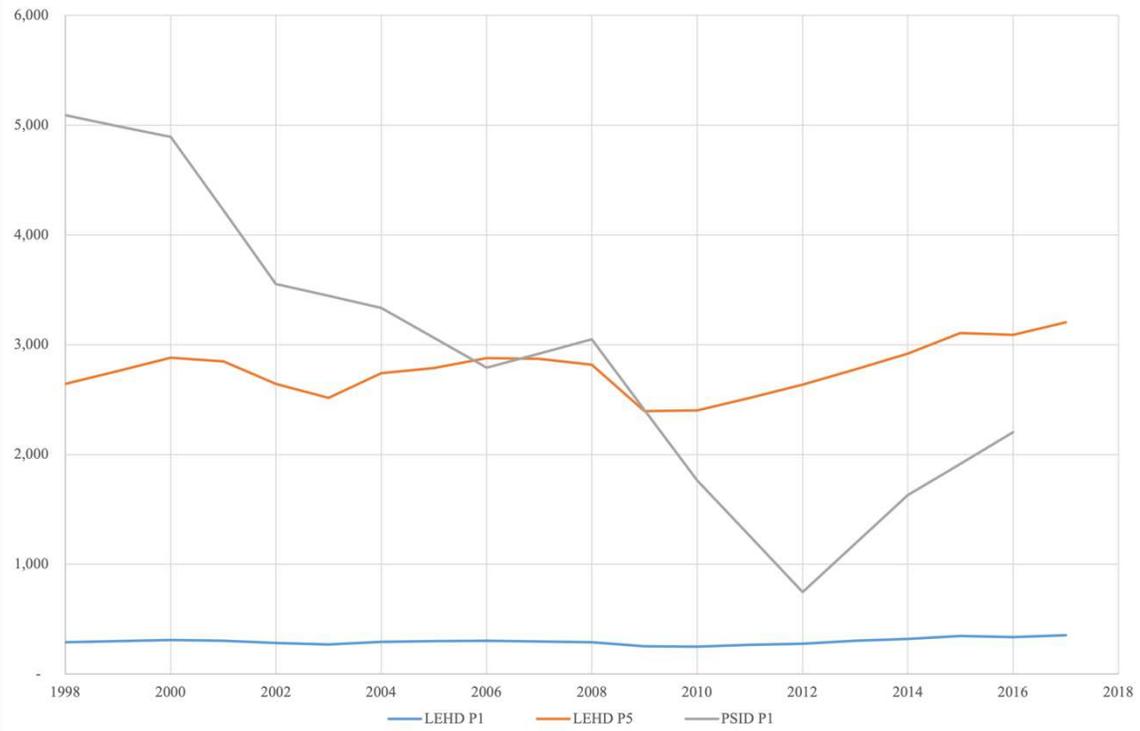

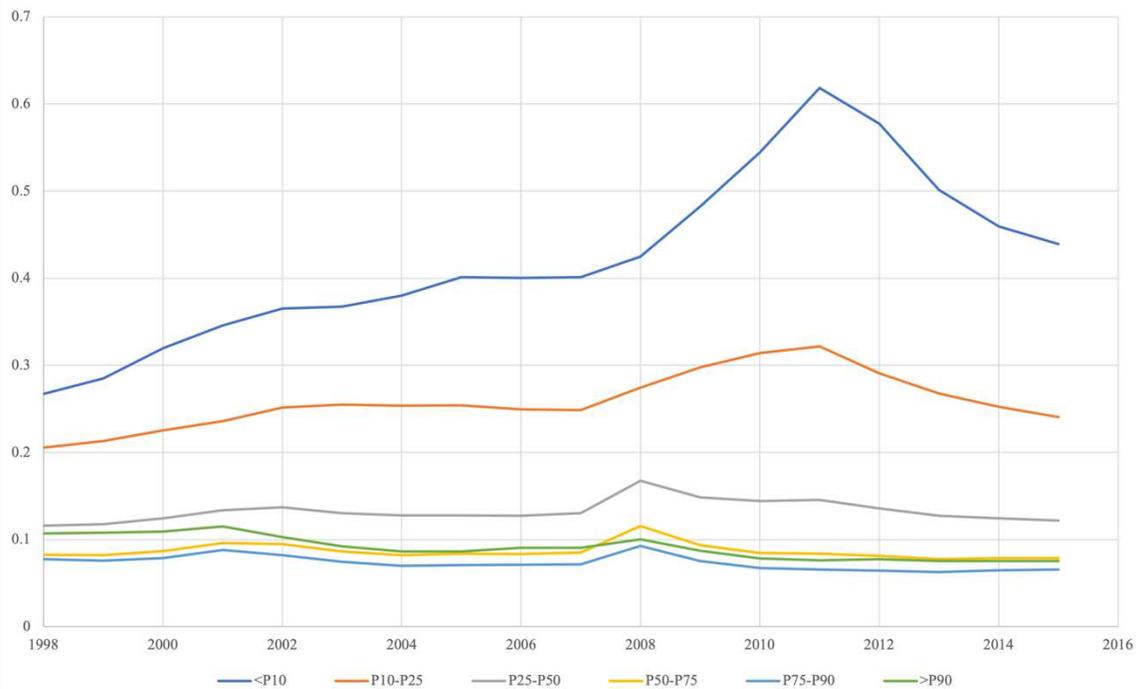

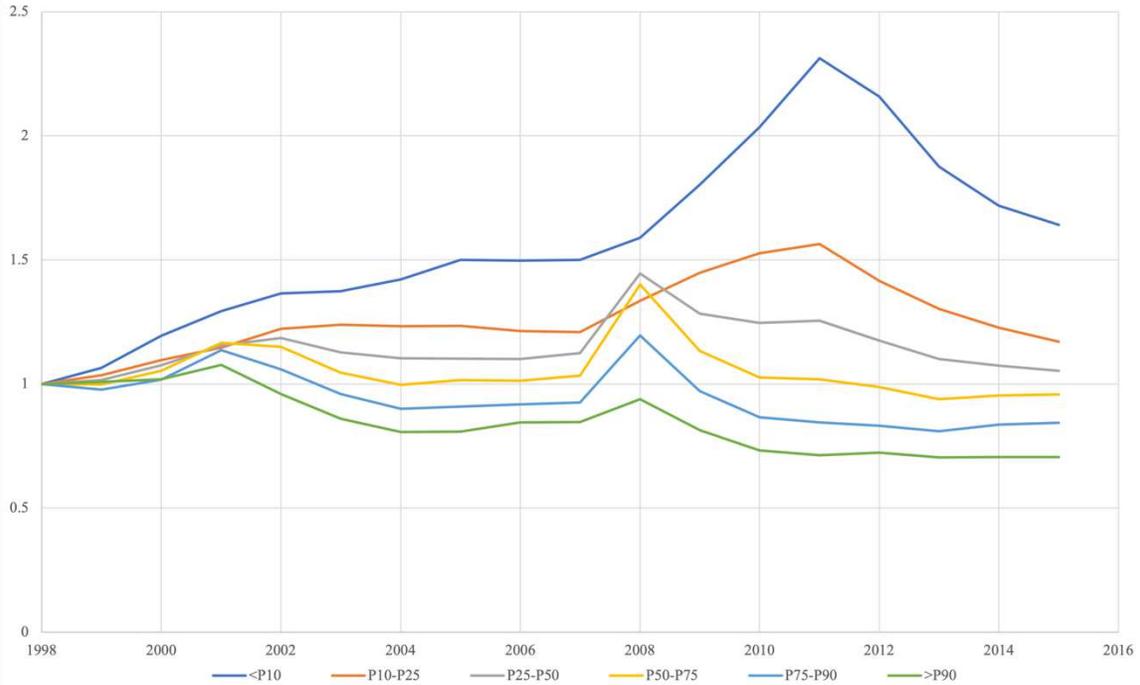

Figure G3 - LEHD Earnings Volatility by Initial Year Earnings (Relative Within Bin Change)

Notes: Real annual earnings (2010 PCE) at all jobs for males age 25-59. Vertical axis shows the variance of the arc percent change (APC, PSID P1 trim) divided by the variance of the arc percent change in 1998. Horizontal axis shows the initial year of the year-pair. Each worker must have reported earnings in both the initial and the subsequent year. Initial year earnings are classified into one of six mutually exclusive by year earnings bins. Percentiles are calculated separately each year; P10~11K, P25~24K, P50~43K, P75~68K, and P90~104K.

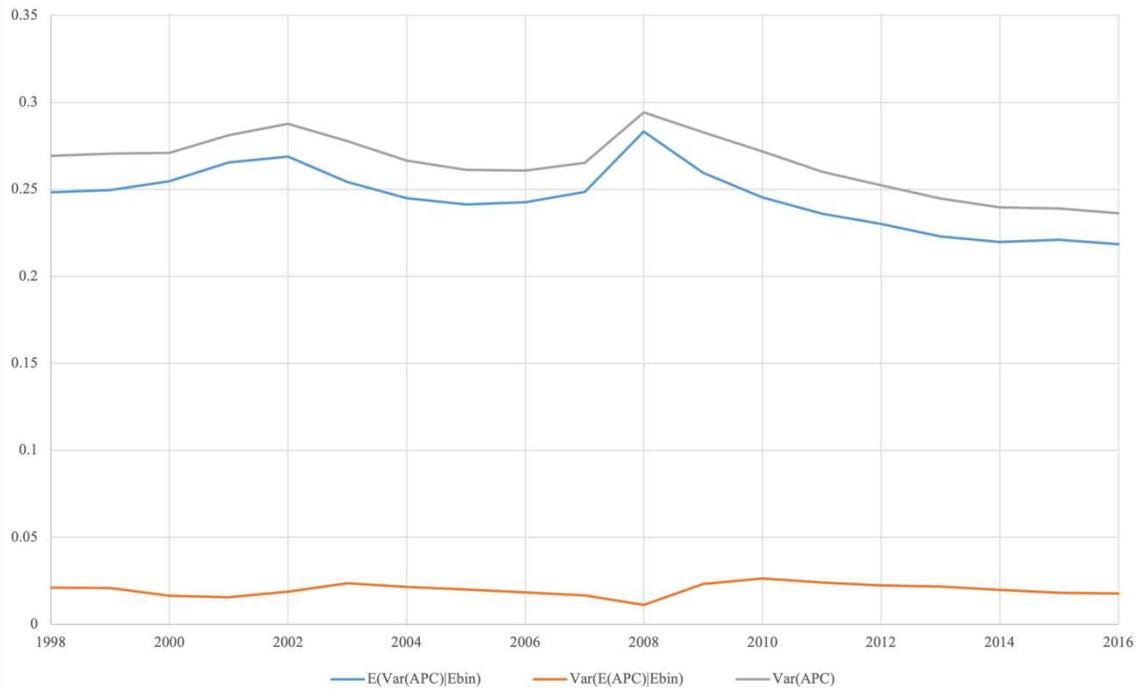

Figure G4 - Earnings Bin Variance Decomposition (APC, No Trim)

Notes: Real annual earnings (2010 PCE) at all jobs for males age 25-59. Vertical axis shows the variance of the arc percent change (APC, no trim). Horizontal axis shows the initial year of the year-pair. Each worker must have reported earnings in both the initial and the subsequent year. Initial year earnings are classified into one of six mutually exclusive by year earnings bins. The decomposition shows two variance components: the weighted average or expectation of within bin volatility and the variance of the across bin APC means. The two components sum to the total, Var(APC).

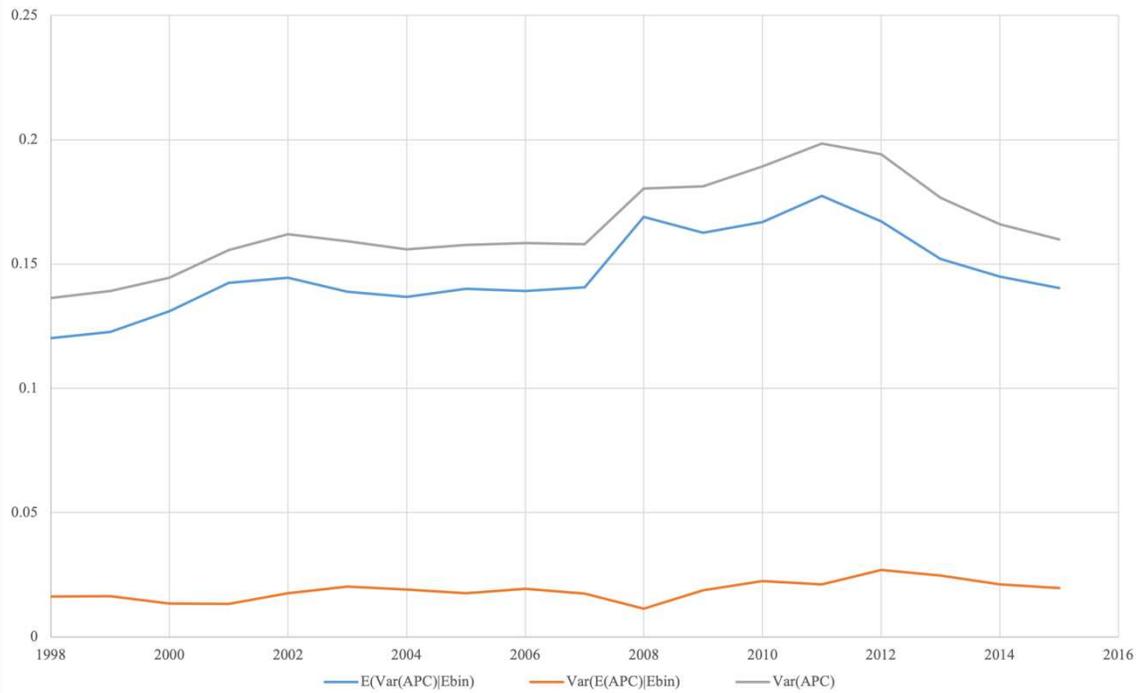

Figure G5 - Earnings Bin Variance Decompostion (APC, PSID P1 Trim)

Notes: Real annual earnings (2010 PCE) at all jobs for males age 25-59. Vertical axis shows the variance of the arc percent change (APC, PSID P1 Trim). Horizontal axis shows the initial year of the year-pair. Each worker must have reported earnings in both the initial and the subsequent year. Initial year earnings are classified into one of six mutually exclusive by year earnings bins. The decomposition shows two variance components: the weighted average or expectation of within bin volatility and the variance of the across bin APC means. The two components sum to the total, Var(APC).